%% file: paper.tex
\documentclass[sigconf]{acmart}

\usepackage{multirow}
\usepackage{appendix}
\usepackage{subfig}
\usepackage{graphicx}
\usepackage{grffile}

\newcounter{subcopyrightbox@save}
\usepackage[font=bf]{caption}
\usepackage{color, url}
\usepackage{xspace} 

\usepackage{amsmath}

\usepackage{mathrsfs}
\usepackage{epstopdf}
\usepackage{balance}
\usepackage{algorithm}
\usepackage{algorithmic}

\newcommand{\argmax}{\operatornamewithlimits{argmax}}
\newcommand{\argmin}{\operatornamewithlimits{argmin}}
\allowdisplaybreaks

\newcommand{\myparatight}[1]{\smallskip\noindent{\bf {#1}:}~}

\newcommand{\sample}{\mathbf{x}}

\newcommand{\confscore}{\mathbf{s}}

\newcommand{\noise}{\mathbf{n}}

\graphicspath{ {../fig/} }

\setcopyright{acmcopyright} 

\begin{document}

\fancyhead{}

\copyrightyear{2019} 
\acmYear{2019} 
\acmConference[CCS '19]{2019 ACM SIGSAC Conference on Computer and Communications Security}{November 11--15, 2019}{London, United Kingdom}
\acmBooktitle{2019 ACM SIGSAC Conference on Computer and Communications Security (CCS '19), November 11--15, 2019, London, United Kingdom}
\acmPrice{15.00}
\acmDOI{10.1145/3319535.3363201}
\acmISBN{978-1-4503-6747-9/19/11}
\settopmatter{printacmref=true}


\title{MemGuard: Defending against Black-Box Membership Inference Attacks via Adversarial Examples}

\author{Jinyuan Jia}
\affiliation{%
  \institution{ECE Department \\ Duke University}
}
\email{jinyuan.jia@duke.edu}

\author{Ahmed Salem}
\affiliation{%
  \institution{CISPA Helmholtz Center for Information Security}
}
\email{ahmed.salem@cispa.saarland}

\author{Michael Backes}
\affiliation{%
  \institution{CISPA Helmholtz Center for Information Security}
}
\email{backes@cispa.saarland}

\author{Yang Zhang}
\affiliation{%
  \institution{CISPA Helmholtz Center for Information Security}
}
\email{zhang@cispa.saarland}

\author{Neil Zhenqiang Gong}
\affiliation{%
  \institution{ECE Department \\ Duke University}
}
\email{neil.gong@duke.edu}

\input{abstract}

\ccsdesc[500]{Security and privacy~}
\ccsdesc[500]{Computing methodologies~Machine learning}

\keywords{Membership inference attacks; adversarial examples; privacy-preserving machine learning}

\maketitle
\input{intro}

\input{related}

\input{problem}

\input{method}

\input{exp}

\input{discussion}
\input{conclusion}

\begin{acks}
We thank the anonymous reviewers for insightful reviews. We would like to thank Hao Chen (University of California, Davis) for discussions. 
This work was partially supported by NSF grant No. 1937786. 
\end{acks}

\bibliographystyle{ACM-Reference-Format}
\bibliography{normal_generated}

\input{appendix}

\end{document}

%% file: abstract.tex
\begin{abstract}
In a membership inference attack, an attacker aims to infer whether a data sample is in a target classifier's training dataset or not. 
Specifically, given a black-box access to the target classifier, the attacker trains a binary classifier, which takes a data sample's confidence score vector predicted by the target classifier as an input and predicts the data sample to be a member or non-member of the target classifier's training dataset. 
Membership inference attacks pose severe privacy and security threats to the training dataset. Most existing defenses leverage differential privacy when training the target classifier or regularize the training process of the target classifier. These defenses suffer from two key limitations: 1) they do not have formal utility-loss guarantees of the confidence score vectors, and 2) they achieve suboptimal privacy-utility tradeoffs.

In this work, we propose \emph{MemGuard}, the first defense with formal utility-loss guarantees against black-box membership inference attacks. Instead of tampering the training process of the target classifier, MemGuard adds noise to each confidence score vector predicted by the target classifier.  Our key observation is that attacker uses a classifier to predict member or non-member and classifier is vulnerable to \emph{adversarial examples}. Based on the observation, we propose to add a carefully crafted noise vector to a confidence score vector to turn it into an adversarial example that misleads the attacker's classifier. Specifically, MemGuard works in two phases. In Phase I, MemGuard finds a carefully crafted noise vector that can turn a confidence score vector into an adversarial example, which is likely to mislead the  attacker's classifier to make a random guessing at member or non-member. We find such carefully crafted noise vector via a new method that we design to incorporate the unique utility-loss constraints on the noise vector. In Phase II, 
MemGuard adds the noise vector to the confidence score vector with a certain probability, which is selected to satisfy a given utility-loss budget on the confidence score vector. Our experimental results on three datasets show that MemGuard can effectively defend against membership inference attacks and achieve better privacy-utility tradeoffs than existing defenses. Our work is the first one to show that adversarial examples can be used as defensive mechanisms to defend against membership inference attacks. 

\end{abstract}

%% file: intro.tex
\section{Introduction}
\label{section:introduction}

Machine learning (ML) is transforming many aspects of our society. We consider a model provider deploys an ML classifier (called \emph{target classifier}) as a black-box software or service, which returns a \emph{confidence score vector} for a query data sample from a user. The confidence score vector is a probability distribution over the possible labels and the label of the query data sample is predicted as the one that has the largest confidence score. Multiple studies have shown that such black-box ML classifier is vulnerable to \emph{membership inference attacks}~\cite{SSSS17,NSH19,SZHBFB19,SSM19}.  
Specifically, an attacker trains a binary classifier, which takes a data sample's confidence score vector predicted by the target classifier as an input and predicts whether the data sample is a \emph{member} or \emph{non-member} of the target classifier's training dataset. Membership inference attacks pose severe privacy and security threats to ML. In particular, in application scenarios where the training dataset is sensitive (e.g., biomedical records and location traces), successful membership inference  leads to severe privacy violations. For instance, if an attacker knows her victim's data is used to train a medical diagnosis classifier,
then the attacker can directly infer the victim's health status. Beyond privacy, membership inference also damages the model provider's intellectual property of the training dataset as collecting and labeling the training dataset may require lots of resources.

Therefore, defending against membership inference attacks is an urgent research problem and multiple defenses~\cite{SSSS17,NSH18,SZHBFB19} have been explored.  A major reason why membership inference attacks succeed is that the target classifier is overfitted. As a result, the confidence score vectors predicted by the target classifier are distinguishable for members and non-members of the training dataset. Therefore, state-of-the-art defenses~\cite{SSSS17,NSH18,SZHBFB19} essentially regularize the training process of the target classifier to reduce overfitting and the gaps of the confidence score vectors between members and non-members of the training dataset. For instance, $L_2$ regularization~\cite{SSSS17}, min-max game based adversarial regularization~\cite{NSH18}, and dropout~\cite{SZHBFB19} have been explored to regularize the target classifier. Another line of defenses~\cite{CMS11,KST12,INSTTW19,SCS13,BST14,WYX17,ACGMMTZ16,YLPGT19} leverage differential privacy~\cite{DMNS06} when training the target classifier.
Since tampering the training process has no guarantees on the confidence score vectors, these defenses have no formal utility-loss guarantees on the confidence score vectors. Moreover, these defenses achieve suboptimal tradeoffs between the membership privacy of the training dataset and utility loss of the confidence score vectors. For instance, Jayaraman and Evans~\cite{JE14} found that existing differentially private machine learning methods rarely offer acceptable privacy-utility tradeoffs for complex models. 

\myparatight{Our work} In this work, we propose \emph{MemGuard}, the first defense with formal utility-loss guarantees against membership inference attacks under the black-box setting.  Instead of tampering the training process of the target classifier, MemGuard randomly adds noise to the confidence score vector predicted by the target classifier for any query data sample. 
MemGuard can be applied to an existing target classifier without retraining it. Given a query data sample's confidence score vector, MemGuard aims to achieve two goals: 1) the attacker's classifier is inaccurate at inferring member or non-member for the query data sample after adding noise to the confidence score vector, and 2) the utility loss of the confidence score vector is bounded. Specifically, the noise should not change the predicted label of the query data sample, since even 1\% loss of the label accuracy may be intolerable in some critical applications such as finance and healthcare. Moreover, the confidence score distortion introduced by the noise should be bounded by a budget since a confidence score vector intends to tell a user more information beyond the predicted label. We formulate achieving the two goals as solving an optimization problem. However, it is computationally challenging to solve the optimization problem as the noise space is large. To address the challenge, we propose a two-phase framework to approximately solve the problem. 

We observe that an attacker uses an ML classifier to predict member or non-member and classifier can be misled by \emph{adversarial examples}~\cite{CW17,PMJFCS16,PMGJCS17,SZSBEGF13,PMSW18,PMG16,KGB16,GSS15}. Therefore,  in Phase I, MemGuard finds a carefully crafted noise vector that can turn the confidence score vector into an adversarial example. Specifically, MemGuard aims to find a noise vector such that the attacker's classifier is likely to make a random guessing at inferring member or non-member based on the noisy confidence score vector. Since the defender does not know the attacker's classifier as there are many choices, the defender itself trains a classifier for membership inference and crafts the noise vector based on its own classifier. Due to \emph{transferability}~\cite{SZSBEGF13,KGB16,LCLS16,PMG16} of adversarial examples, the noise vector that misleads the defender's classifier is likely to also mislead the attacker's classifier. The adversarial machine learning community has developed many algorithms (e.g.,~\cite{CW17,PMJFCS16,GSS15,MMSTV18,KGB16,MFF16,TKPGBM17,MFFF17}) to find adversarial noise/examples. However, these algorithms are insufficient for our problem because they did not consider the unique constraints on utility loss of the confidence score vector. Specifically, the noisy confidence score vector should not change the predicted label of the query data sample and should still be a probability distribution. To address this challenge, we design a new algorithm to find a small noise vector that satisfies the utility-loss constraints. 

In Phase II, MemGuard adds the noise vector found in Phase I to the true confidence score vector with a certain probability. The probability is selected such that the expected confidence score distortion is bounded by the budget and the defender's classifier is most likely to make random guessing at inferring member or non-member.  Formally, we formulate finding this probability as solving an optimization problem and derive an analytical solution for the optimization problem. 

We evaluate MemGuard and compare it with state-of-the-art defenses~\cite{SSSS17,NSH18,SZHBFB19,ACGMMTZ16} on three real-world datasets. Our empirical results show that MemGuard can effectively defend against state-of-the-art black-box membership inference attacks~\cite{NSH19,SZHBFB19}. In particular, as MemGuard is allowed to add larger noise (we measure the magnitude of the noise using its $L_1$-norm), the inference accuracies of all evaluated membership inference attacks become smaller. Moreover, MemGuard achieves better privacy-utility tradeoffs than state-of-the-art defenses. Specifically, given the same average confidence score distortion, MemGuard reduces the attacker's inference accuracy at inferring member/non-members by the most. 

In summary, our key contributions are as follows:
\begin{itemize}
\item We propose MemGuard, the first defense with formal utility-loss guarantees against membership inference attacks under the black-box setting.  
\item We propose a new algorithm to find a noise vector that satisfies the unique utility-loss constraints in Phase I of MemGuard. Moreover, in Phase II, we derive an analytical solution of the probability with which MemGuard adds the noise vector to the confidence score vector. 
\item We evaluate MemGuard on three real-world datasets. Our results show that MemGuard is effective and outperforms existing defenses. 
\end{itemize}

%% file: related.tex
\section{Related Work}
\label{related}

\subsection{Membership Inference}

\myparatight{Membership inference attacks}
The goal of membership inference is to determine whether a certain data sample is inside a dataset.
Homer et al.~\cite{HSRDTMPSNC08} proposed the first membership inference attack in the biomedical setting,
in particular on genomic data.  Specifically, they showed that an attacker can compare a user's genomic data with the summary statistics
of the target database, such as mean and standard deviation, 
to determine the presence of the user in the database.
The comparison can be done by using statistical testing methods such as log-likelihood ratio test.
Later, several works performed similar membership inference attacks against other types of biomedical data such as MicroRNA~\cite{BBHM16} 
and DNA methylation~\cite{HZHBTWB19}.
Recently, Pyrgelis et al.~\cite{PTC18,PTC19} further showed that 
membership inference can also be performed effectively against location databases. In particular, they showed that an attacker can infer whether a user's location dataset was used for computing a given aggregate location dataset. 

\myparatight{Membership inference attacks against ML models}
Shokri et al.~\cite{SSSS17} introduced membership inference in the ML setting. 
The goal here is to determine whether a data sample is in the training dataset of a target black-box ML classifier.
To achieve the goal, the attacker trains binary ML classifiers, which take a data sample's {confidence score vector} predicted by the target classifier as input and infer the data sample to be a member or non-member of the target classifier's training dataset. We call these classifiers \emph{attack classifiers} and they are trained using \emph{shadow classifiers}. Specifically, the attacker is assumed to have a dataset coming from the same distribution as the target classifier's training dataset and the attacker uses the dataset to train {shadow classifiers}, each of which aims to replicate the target classifier. Then, the attacker trains the attack classifiers by using the confidence score vectors predicted by the shadow classifiers for some members and non-members of the shadow classifiers' training datasets.

Salem et al.~\cite{SZHBFB19} recently proposed new membership inference attacks for black-box target classifiers, which relax the assumptions of the attacks proposed by Shokri et al. 
from both model and data angles. For instance, they showed that the attacker can rank the entries in a confidence score vector before feeding it into an attack classifier, which improves the attack effectiveness. Moreover, they showed that it is sufficient for the attacker to train just one shadow classifier. These results 
indicate that membership inference threat is even larger than previously thought.

More recently, Nasr et al.~\cite{NSH19} proposed membership inference attacks against white-box ML models.
For a data sample,
they calculate the corresponding gradients over the white-box target classifier's parameters 
and use these gradients as the data sample's feature for membership inference. Moreover, both Nasr et al.~\cite{NSH19} and Melis et al.~\cite{MSCS19} proposed membership inference attacks against federated learning. 
While most of the previous works concentrated on classification models~\cite{SSSS17,LBG17,LBWBWTGC18,NSH18,YGFJ18,SZHBFB19,NSH19},
Hayes et al.~\cite{HMDC19} studied membership inference against generative models, 
in particular generative adversarial networks (GANs)~\cite{GPMXWOCB14}.
They designed attacks for both white- and black-box settings.
Their results showed that generative models are also vulnerable to membership inference.

\myparatight{Defense mechanisms against membership inference}
Multiple defense mechanisms have been proposed to mitigate the threat of membership inference in the ML setting.
We summarize them as the following.

 {\bf $L_2$-Regularizer~\cite{SSSS17}.} Overfitting, i.e., ML classifiers are more confident when facing data samples they are trained on (members)
    than others, is one major reason why membership inference is effective. Therefore, to defend against membership inference, people have explored to reduce overfitting using regularization. 
    For instance, Shokri et al.~\cite{SSSS17} explored using conventional $L_2$ regularizer when training the target classifier.
 
{\bf Min-Max Game~\cite{NSH18}.} Nasr et al.~\cite{NSH18} proposed a min-max game-theoretic method to train a target classifier. Specifically, the method formulates a min-max optimization problem that aims to minimize the target classifier's prediction loss while maximizing the membership privacy. This formulation is equivalent to adding a new regularization term called \emph{adversarial regularization} to the loss function of the target classifier.

{\bf Dropout~\cite{SZHBFB19}.} Dropout is a recently proposed technique to regularize neural networks~\cite{SHKSS14}.  Salem et al.~\cite{SZHBFB19} explored using dropout to mitigate membership inference attacks. Roughly speaking, dropout drops a neuron with a certain probability in each iteration of training a neural network. 

{\bf Model Stacking~\cite{SZHBFB19}.}
    Model stacking is a classical ensemble method which combines multiple weak classifiers' results as a strong one. Salem et al.~\cite{SZHBFB19} explored using model stacking to mitigate membership inference attacks. 
    Specifically, the target classifier consists of  three classifiers organized into a two-level tree structure. 
    The first two classifiers on the bottom of the tree take
the original data samples as input, while the third one's input is the outputs of the first two classifiers.
        The three classifiers are trained using disjoint sets of data samples, 
    which reduces the chance for the target classifier to remember any specific data sample, thus preventing overfitting.
    
 {\bf Differential privacy.}  Differential privacy~\cite{DMNS06} is a classical method for privacy-preserving machine learning. 
Most differential privacy based defenses add noise to the objective function that is used to learn a model~\cite{CMS11,KST12,INSTTW19}, or the gradient in each iteration of gradient descent or stochastic gradient descent that is used to minimize the objective function~\cite{SCS13,BST14,WYX17,ACGMMTZ16,YLPGT19}. Shokri and Shmatikov~\cite{SS15} designed a differential privacy method for collaborative learning of deep neural networks.

{\bf Limitations.} Existing defenses suffer from two key limitations: 1) they do not have formal utility loss guarantee of the confidence score vector; and 2) they achieve suboptimal privacy-utility tradeoffs. Our defense addresses these two limitations. For instance, as we will show in experiments, with the same utility loss of the confidence score vector (e.g., the same $L_1$-norm distortion of the confidence score vector), our defense reduces the attack classifier's 
accuracy at inferring members/non-members to a larger extent than existing defenses.

\myparatight{Other privacy/confidentiality attacks against ML}
There exist multiple other types of privacy/confidentiality attacks against ML models~\cite{FLJLPR14,FJR15,AFMSVV13,GWYGB18,MSCS19,TZJRR16,WG18,OASF18,SBBFZ19}.
Fredrikson et al.~\cite{FLJLPR14,FJR15} proposed \emph{model inversion attacks}. For instance, 
they can infer the missing values of an input feature vector
by leveraging a classifier's prediction on the input feature vector. Several works~\cite{AFMSVV13,GWYGB18,MSCS19} studied \emph{property inference attacks}, which aim to infer a certain property (e.g., the fraction of male and female users) of a target classifier's training dataset.
Tram{\`{e}}r et al.~\cite{TZJRR16}
proposed \emph{model stealing attacks}.
They designed different techniques tailored to different ML models
aiming at stealing the parameters of the target models.
Another line of works studied \emph{hyperparameter stealing attacks}~\cite{WG18,OASF18}, which aim to steal the hyperparameters such as the neural network architecture and the hyperparameter that balances between the loss function and the regularization term.

\subsection{Adversarial Examples}
\label{adversarialexample}

Given a classifier and an example, we can add carefully crafted noise to the example such that the classifier predicts its label as we desire. The example with carefully crafted noise is called an \emph{adversarial example}. Our MemGuard adds carefully crafted noise to a confidence score vector to turn it into an adversarial example, which is likely to mislead the attack classifier to make a random guessing at member or non-member. The adversarial machine learning community has developed many algorithms (e.g.,~\cite{CW17,PMJFCS16,GSS15,MMSTV18,KGB16,MFF16,TKPGBM17,MFFF17}) to find adversarial examples. However, these algorithms are insufficient to our problem because they did not consider the utility-loss constraints on the confidence score vectors. We address these challenges via designing a new algorithm to find adversarial examples.

Since our defense leverages adversarial examples to mislead the attacker's attack classifier, an adaptive attacker can leverage a classifier that is more robust against adversarial examples as the attack classifier. Although different methods (e.g., adversarial training~\cite{GSS15,TKPGBM17,MMSTV18}, defensive distillation~\cite{PMWJS16}, Region-based Classification~\cite{CG17}, MagNet~\cite{MC17}, and Feature Squeezing~\cite{XEQ18}) have been explored to make classifiers robust against adversarial examples, it is still considered an open challenge to design such robust classifiers. Nevertheless, in our experiments, we will consider the attacker uses adversarial training to train its attack classifier, as adversarial training was considered to be the most empirically robust method against adversarial examples so far~\cite{ACW18}.

%% file: problem.tex
\section{Problem Formulation}
\label{section:problem}
In our problem formulation, we have three parties, i.e., \emph{model provider}, \emph{attacker}, and \emph{defender}. \autoref{table:notations} shows some important notations used in this paper. 

\begin{table}[!t]
\centering
\caption{Notations}
\label{table:notations}
\addtolength{\tabcolsep}{-2pt}
\begin{tabular}{c|l} 
\toprule
Notation & Description\\
\midrule
$\sample$ & A data sample\\
$\confscore$ & A true confidence score vector\\
$\mathbf{s}^{\prime}$ & A noisy confidence score vector\\
$\noise$ & A noise vector\\
$f$ & Decision function of the target classifier\\
$\mathbf{z}$ & Logits of the target classifier\\
$C$ & Attacker's attack classifier for membership inference\\
$g$ & Decision function of defender's defense classifier\\
$h$& Logits of the defender's defense classifier\\
$\mathcal{M}$ & Randomized noise addition mechanism\\
$\epsilon$ & Confidence score distortion budget\\
\bottomrule
\end{tabular}
\end{table}

\subsection{Model Provider}
We assume a model provider has a proprietary training dataset (e.g., healthcare dataset, location dataset). The model provider trains a machine learning classifier using the proprietary training dataset. Then, the model provider deploys the classifier as a cloud service or a client-side AI software product (e.g., a mobile or IoT app), so other users can leverage the classifier to make predictions for their own data samples. In particular, we consider the deployed classifier returns a  {confidence score vector} for a query data sample. 
Formally, we have:
\[
f: \sample \mapsto \confscore,
\]
where $f$, $\sample$, and $\confscore$ represent the classifier's decision function, the query data sample, and the confidence score vector, respectively. The confidence score vector essentially is the predicted posterior probability distribution of the label of the query data sample, i.e., $s_j$ is the predicted posterior probability that the query data sample has label $j$. The label of the query data sample is predicted to be the one that has the largest confidence score, i.e., the label is predicted as $\argmax_j \{s_j\}$. For convenience, we call the model provider's classifier \emph{target classifier}. Moreover, we consider the target classifier is neural network in this work.

\subsection{Attacker}
An attacker aims to infer the proprietary training dataset of the model provider. Specifically, we consider the attacker only has \emph{black-box} access to the target classifier, i.e., the attacker can send query data samples to the target classifier and obtain their confidence score vectors predicted by the target classifier. The attacker leverages black-box \emph{membership inference attacks}~\cite{LBWBWTGC18,NSH18,SSSS17,SZHBFB19} to infer the members of the target classifier's training dataset. Roughly speaking, in membership inference attacks, the attacker trains a binary classifier, which takes a query data sample's confidence score vector as input and predicts whether the query data sample is in the target classifier's training dataset or not. Formally, we have:
\[
C: \confscore \mapsto \{0, 1\},
\]
where $C$ is the attacker's binary classifier, $\confscore$ is the confidence score vector predicted by the target classifier for the query data sample $\sample$, 0 indicates that the query data sample $\sample$ is not a member of the target classifier's training dataset, and 1 indicates that the query data sample $\sample$ is a member of the target classifier's training dataset. For convenience, we call the attacker's binary classifier $C$ \emph{attack classifier}.  We will discuss more details about how the attacker could train its attack classifier in Section~\ref{evaluation_section}. Note that, to consider strong attacks, we assume the attacker  knows our defense mechanism, but the defender does not know the attack classifier since the attacker has many choices for the attack classifier. 

\subsection{Defender}
The defender aims to defend against black-box membership inference attacks. The defender could be the model provider itself or a trusted third party. For any query data sample from any user, the target classifier predicts its confidence score vector and the defender adds a \emph{noise vector} to the confidence score vector before returning it to the user. Formally, we have:
\[
\confscore'=\confscore+\noise,
\]
where $\confscore$ is the true confidence score vector predicted by the target classifier for a query data sample,  $\noise$ is the noise vector added by the defender, and $\confscore'$ is the noisy confidence score vector that is returned to a user. Therefore, an attacker only has access to the noisy confidence score vectors. 
The defender aims to add noise to achieve the following two goals:
\begin{itemize}
\item {\bf Goal I.} The attacker's attack classifier is inaccurate at inferring the members/non-members of the target classifier's training dataset, i.e., protecting the privacy of the training dataset.
\item {\bf Goal II.} The utility loss of the confidence score vector is bounded. 
\end{itemize}

However, achieving these two goals faces several challenges which we discuss next. 

\myparatight{Achieving Goal I} The first challenge to achieve Goal I is that the defender does not know the attacker's attack classifier.  
To address the challenge, the defender itself trains a binary classifier to perform membership inference and adds noise vectors to the confidence score vectors such that its own classifier is inaccurate at inferring members/non-members. In particular, the defender's classifier takes a confidence score vector as input and predicts member or non-member for the corresponding data sample. We call the defender's binary classifier \emph{defense classifier} and denote its decision function as $g$. Moreover, we consider the decision function $g(\confscore)$ represents the probability that the corresponding data sample, whose confidence score vector predicted by the target classifier is $\confscore$,  is a member of  the target classifier's training dataset. In particular, we consider the defender trains a neural network classifier, whose output layer has one neuron with sigmoid activation function. For such classifier, the decision function's output (i.e., the output of the neuron in the output layer) represents probability of being a member.  Formally, we have:
\[
g: \confscore \mapsto [0, 1].
\]
The defense classifier predicts a data sample to be member of the target classifier's training dataset if and only if $g(\confscore)>0.5$.

To make the defense classifier inaccurate, one method is to add a noise vector to a true confidence score vector such that  the defense classifier makes an incorrect prediction. Specifically, if the defense classifier predicts member (or non-member) for the true confidence score vector, then the defender adds a noise vector such that the defense classifier predicts non-member (or member) for the noisy confidence score vector. However, when an attacker knows the defense mechanism, the attacker can easily adapt its attack to achieve a high accuracy. In particular, the attacker predicts member (or non-member) when its attack classifier predicts non-member (or member) for a data sample.  Another method is to add noise vectors such that the defense classifier always predicts member (or non-member) for the noisy confidence score vectors. However, for some true confidence score vectors, such method may need noise that violates the utility-loss constraints of the confidence score vectors (we will discuss utility-loss constraints later in this section). 

{\bf Randomized noise addition mechanism.} Therefore, we consider the defender adopts a \emph{randomized noise addition mechanism} denoted as $\mathcal{M}$. Specifically, given a true confidence score vector $\confscore$, the defender samples a noise vector  $\noise$ from the space of possible noise vectors with a probability $\mathcal{M}(\noise|\confscore)$ and adds it to the true confidence score vector.  
Since random noise is added to a true confidence score vector, the decision function $g$ outputs a random probability of being member. We consider the defender's goal is to make the expectation of the probability of being member predicted by $g$ close to 0.5. In other words, the defender's goal is to add random noise such that the defense classifier randomly guesses member or non-member for a data sample on average. Formally, the defender aims to find a randomized noise addition mechanism $\mathcal{M}$ such that $|E_{\mathcal{M}}(g(\confscore + \noise))-0.5|$ is minimized.

\myparatight{Achieving Goal II} The key challenge to achieve Goal II is how to quantify the utility loss of the confidence score vector. To address the challenge, we introduce two utility-loss metrics. 

 {\bf Label loss.}
Our first metric concentrates on the query data sample's label predicted by the target classifier. Recall that the label of a query data sample is predicted as the one that has the largest confidence score. If the true confidence score vector and the noisy confidence score vector predict the same label for a query data sample, then the \emph{label loss} is 0 for the query data sample, otherwise the label loss is 1 for the query data sample. The overall label loss of a defense mechanism is the label loss averaged over all query data samples. In some critical applications such as finance and healthcare, even 1\% of label loss may be intolerable. In this work, we aim to achieve 0 label loss, i.e., our noise does not change the predicted label  for any query data sample. Formally, we aim to achieve   $\argmax_j \{s_j\} = \argmax_j \{s_j + n_j\}$, where $\argmax_j \{s_j\}$ and $\argmax_j \{s_j + n_j\}$ are the labels predicted based on the true and noisy confidence score vectors, respectively.

{\bf Confidence score distortion.} The confidence score vector for a query data sample tells the user more information about the data sample's label beyond the predicted label. Therefore, the added noise should not substantially distort the confidence score vector. First, the noisy confidence score vector should still be a probability distribution. Formally, we have $s_j+n_j\geq 0$ for $\forall j$ and $\sum_j (s_j+n_j)=1$. Second, the distance  $d(\confscore, \confscore+\noise)$ between the true confidence score vector and the noisy confidence score vector should be small. In particular, we consider the model provider specifies a confidence score distortion budget called $\epsilon$, which indicates the upper bound of the expected confidence score distortion that the model provider can tolerate. Formally, we aim to achieve  $E_{\mathcal{M}}(d(\confscore,\confscore+\noise))\leq \epsilon$. While any distance metric can be used to measure the distortion, we consider $L_1$-norm of the noise vector as the distance metric, i.e., $d(\confscore,\confscore+\noise)=||\noise||_1$. We adopt $L_1$-norm of the noise vector because it is easy to interpret. Specifically, the $L_1$-norm of the noise vector is simply the sum of the absolute value of its entries.

\myparatight{Membership inference attack defense problem} After quantifying  Goal I and Goal II, we can formally define our problem of defending against membership inference attacks.

\begin{definition}[Membership-Inference-Attack Defense Problem]
    Given the decision function $g$ of the defense classifier, a confidence score distortion budget $\epsilon$, a true confidence score vector $\confscore$, the defender aims to find a randomized noise addition mechanism $\mathcal{M}^{\ast}$ via solving the following optimization problem: 
\begin{align}
\label{originalproblem}
    \mathcal{M}^{\ast}&=\argmin_{\mathcal{M}} |E_{\mathcal{M}}(g(\confscore+\mathbf{n}))-0.5| \\
    \label{originalproblem-c1}
    \text{subject to: } & \argmax_j \{s_j+{n}_j\} =\argmax_j \{s_j\}  \\
        \label{originalproblem-c2}
    &E_{\mathcal{M}}(d(\confscore,\confscore+\mathbf{n}))\leq \epsilon \\
        \label{originalproblem-c3}
    &s_j+{n}_j \geq 0,\forall j \\
        \label{originalproblem-c4}
    &\sum_j s_j + {n}_j=1,  
\end{align}
where the objective function of the optimization problem is to achieve  Goal I and the constraints are to achieve  Goal II. Specifically, the first constraint means that the added noise does not change the predicted label of the query data sample; the second constraint means that the confidence score distortion is bounded by the budget $\epsilon$; and the last two constraints mean that the noisy confidence score vector is still a probability distribution. Note that the last constraint is equivalent to $\sum_j  {n}_j=0$ since $\sum_j  {s}_j=1$. Moreover, we adopt $L_1$-norm of the noise vector to measure the confidence score distortion, i.e., $d(\confscore,\confscore+\mathbf{n})=||\noise||_1$. 
\label{optimization_original}
\end{definition}

%% file: method.tex
\section{Our M\lowercase{em}G\lowercase{uard}}

\subsection{Overview}
Finding the randomized noise addition mechanism is to solve the optimization problem in~\autoref{originalproblem}. We consider two scenarios depending on whether $g(\confscore)$ is 0.5 or not. 

\myparatight{Scenario I} In this scenario, $g(\mathbf{s})=0.5$. For such scenario, it is easy to solve the optimization problem in~\autoref{originalproblem}. Specifically, the mechanism that adds the noise vector $\mathbf{0}$ with probability 1 is the optimal randomized noise addition mechanism, with which the objective function has a value of 0.

\myparatight{Scenario II}
In this scenario, $g(\confscore)$ is not 0.5. 
The major challenge to solve the optimization problem in this scenario is that the randomized noise addition mechanism is a probability distribution over the continuous noise space for a given true confidence score vector. The noise space consists of the noise vectors that satisfy the four constraints of the optimization problem.  As a result, it is challenging to represent the probability distribution and solve the optimization problem. To address the challenge, we observe that the noise space can be divided into two groups depending on the output of the defense classifier's decision function $g$. 
Specifically, for noise vectors in one group, if we add any of them to the true confidence score vector, then the decision function $g$ outputs 0.5 as the probability of being member. For noise vectors in the other group, if we add any of them to the true confidence score vector, then the decision function $g$ outputs a probability of being member that is not 0.5. 

Based on this observation, we propose a \emph{two-phase framework} to approximately solve the optimization problem. Specifically, in Phase I, for each noise group, we find the noise vector with minimum confidence score distortion (i.e., $d(\confscore,\confscore+\mathbf{n})$ is minimized) as a \emph{representative} noise vector for the noise group. We select the noise vector with minimum confidence score distortion in order to minimize the confidence score distortion. Since $g(\confscore)\neq 0.5$, the selected representative noise vector for the second noise group is $\mathbf{0}$. We denote by $\mathbf{r}$ the selected representative noise vector for the first noise group. In Phase II, we assume the randomized noise addition mechanism is a probability distribution over the two representative noise vectors instead of the overall noise space. Specifically, the defender adds the representative noise vector $\mathbf{r}$ to the true confidence score vector with a certain probability and does not add any noise with the remaining probability. 

Next, we introduce our Phase I and Phase II.

\subsection{Phase I: Finding $\mathbf{r}$}
\label{sec:phaseI}
\myparatight{Finding $\mathbf{r}$ as solving an optimization problem} Our goal essentially is to find a noise vector $\mathbf{r}$ such that 1) the utility loss of the confidence score vector is minimized and 2) the decision function $g$ outputs 0.5 as the probability of being member when taking the noisy confidence score vector as an input. Formally, we find such noise vector via solving the following optimization problem:

\begin{align}
\label{optimization_p_1}
    \min_{\mathbf{r}}&\ d(\mathbf{s},\mathbf{s}+\mathbf{r})  \\
    \label{prediction_label}
    \text{subject to: } & \argmax_j \{s_j+r_j\} =\argmax_j \{s_j\}  \\
    \label{noise_goal_constraint}
    & g(\mathbf{s}+\mathbf{r})=0.5  \\
    \label{probability_constraint_1}
    &s_j+r_j \geq 0,\forall j \\
    \label{probability_constraint_2}
    &\sum_j r_j=0,  
\end{align}
where $\mathbf{s}$ is the true confidence score vector, the objective function means that the confidence score distortion is minimized, the first constraint means that the noise does not change the predicted label of the query data sample, the second constraint means that the defense classifier's decision function outputs 0.5 (i.e., the defense classifier's prediction is random guessing), and the last two constraints mean that the noisy confidence score vector is still a probability distribution.

Solving the optimization problem in~\autoref{optimization_p_1} can be viewed as finding an \emph{adversarial example} to evade the defense classifier. In particular, $\mathbf{s}$ is a normal example and $\mathbf{s+r}$ is an adversarial example. The adversarial machine learning community has developed many algorithms (e.g.,~\cite{CW17,PMJFCS16,GSS15,MMSTV18,KGB16,MFF16,TKPGBM17,MFFF17}) to find adversarial examples. However, these algorithms are insufficient to our problem because they did not consider the unique challenges of privacy protection. In particular, they did not consider the utility-loss constraints, i.e., the constraints in~\autoref{prediction_label},~\autoref{probability_constraint_1}, and~\autoref{probability_constraint_2}. 

One naive method (we call it \emph{Random}) to address the challenges is to generate a {random} noise vector that satisfies the utility-loss constraints. In particular, we can generate a random vector $\mathbf{r}^{\prime}$ whose entries are non-negative and sum to 1. For instance, we first sample a number ${r'_1}$ from the interval [0,1] uniformly at random as the first entry. Then, we sample a number  ${r'_2}$ from the interval [0, 1-${r'_1}$] uniformly at random as the second entry. We repeat this process until the last entry is 1 minus the sum of the previous entries. Then, we exchange the largest entry of $\mathbf{r}^{\prime}$ to the position $j$ to satisfy the constraint~\ref{prediction_label}. Finally, we treat $\mathbf{r}=\mathbf{r}^{\prime}-\mathbf{s}$ as the noise vector, which is a solution to the optimization problem in~\autoref{optimization_p_1}. However, as we will show in experiments, this Random method achieves suboptimal privacy-utility tradeoffs because the noise vector is not optimized and it is challenging to satisfy the constraint~\autoref{probability_constraint_1}. 
We propose to solve the optimization problem via change of variables and adding the constraints to the objective function.

\myparatight{Eliminating the constraints on probability distribution via change of variables} Since we consider the target classifier to be a neural network, whose output layer is a softmax layer,  the true confidence score vector $\confscore$ is a {softmax function} of some vector $\mathbf{z}$. The vector $\mathbf{z}$ is the output of the neurons in the second-to-last layer of the neural network and is often called \emph{logits} of the neural network. Formally, we have: 
\begin{align}
  \mathbf{s}&=softmax(\mathbf{z}).
\end{align}
Moreover, we model the noisy confidence score vector as follows:
\begin{align}
  \mathbf{s}+\mathbf{r}&=softmax(\mathbf{z}+\mathbf{e}),
\end{align}
where $\mathbf{e}$ is a new vector variable. For any value of $\mathbf{e}$, the noisy confidence score vector $\mathbf{s}+\mathbf{r}$ is a probability distribution, i.e., the constraints in~\autoref{probability_constraint_1} and~\autoref{probability_constraint_2} are satisfied. Therefore, in the optimization problem in~\autoref{optimization_p_1}, we change the true confidence score vector $\mathbf{s}$ as  $softmax(\mathbf{z})$ and change the variable $\mathbf{r}$ as $softmax(\mathbf{z}+\mathbf{e}) - softmax(\mathbf{z})$. Then, we obtain the following optimization problem:

\begin{align}
    \label{l_1_loss_constraints}
    \min_{\mathbf{e}}&\ d(softmax(\mathbf{z}),softmax(\mathbf{z}+\mathbf{e})) \\
    \label{constraints_1}
    \text{subject to: } & \argmax_j \{z_j+e_j\} =\argmax_j \{z_j\} \\
    \label{constraints_2}
    &g(softmax(\mathbf{z}+\mathbf{e}))= 0.5.
\end{align}
After solving $\mathbf{e}$ in the above optimization problem,  
we can obtain the noise vector $\mathbf{r}$ as follows: 
\begin{align}
\label{noise_equation}
    \mathbf{r}=softmax(\mathbf{z}+\mathbf{e})-softmax(\mathbf{z}).
\end{align}

The optimization problem without the constraints on probability distribution is still challenging to solve because the remaining two constraints are highly nonlinear. To address the challenge, we turn the constraints into the objective function.

\myparatight{Turning the constraint in~\autoref{constraints_2} into the objective function} We consider the defender's binary defense classifier is a neural network whose output layer has a single neuron with sigmoid activation function. Therefore, we have:
\begin{align}
 g(softmax(\mathbf{z}+\mathbf{e}))=\frac{1}{1+\exp(-h(softmax(\mathbf{z}+\mathbf{e})))},   
\end{align}
where $h(softmax(\mathbf{z}+\mathbf{e}))$ is the output of the neuron in the second-to-last layer of the defense classifier when the defense classifier takes the noisy confidence score vector $softmax(\mathbf{z}+\mathbf{e})$ as an input. In other words, $h$ is the logit of the defense classifier. 
 $g(softmax(\mathbf{z}+\mathbf{e}))=0.5$ implies $h(softmax(\mathbf{z}+\mathbf{e}))=0$. Therefore, we transform the constraint in~\autoref{constraints_2} to the following loss function:
 \begin{align}
 L_1=|h(softmax(\mathbf{z}+\mathbf{e}))|,   
\end{align}
where $L_1$ is small when $h(softmax(\mathbf{z}+\mathbf{e}))$ is close to 0.

\begin{algorithm}[t]
    \caption{Phase I of MemGuard}
    \begin{algorithmic}[1]
    \REQUIRE $\mathbf{z}$, $max\_iter$, $c_2$, $c_3$, and $\beta$ (learning rate). \\
    \ENSURE $\mathbf{e}$ \\
	\STATE //Predicted label \;
        \STATE $l=\argmax_{j}\{z_j\}$
        \WHILE {$True$} \;
        \STATE //A new iteration to search $c_3$ \;
        \STATE $\mathbf{e}=\mathbf{0}$
        \STATE $\mathbf{e}'=\mathbf{e}$
         \STATE $i=1$
        \WHILE {$i<max\_iter$ and ($\argmax_{j}\{z_j+e_j\}\neq l$ or $h(softmax(\mathbf{z}))\cdot h(softmax(\mathbf{z}+\mathbf{e})) > 0$)} \;
        \label{conditioncheck}
        \STATE //Gradient descent with normalized gradient \;
        \STATE $\mathbf{u}=\frac{\partial L}{\partial \mathbf{e}}$
        \STATE $\mathbf{u}=\mathbf{u}/||\mathbf{u}||_2$
        \STATE $\mathbf{e}=\mathbf{e}-\beta \cdot \mathbf{u}$
        \STATE $i=i+1$
        \ENDWHILE \;
        \STATE //Return the vector in the previous iteration if the predicted label changes or the sign of $h$ does not change in the current iteration
        \IF{$\argmax_{j}\{z_j+e_j\}\neq l$ or $h(softmax(\mathbf{z}))\cdot h(softmax(\mathbf{z}+\mathbf{e})) > 0$} \;
        \RETURN $\mathbf{e}'$
        \ENDIF \;
        \STATE $c_3=10\cdot c_3$
        \ENDWHILE \;
    \end{algorithmic}
    \label{algorithml1}
    \end{algorithm}

\myparatight{Turning the constraint in~\autoref{constraints_1} into the objective function} 
We denote by $l$ the predicted label for the query data sample, i.e., $l=argmax_{j}\{s_j\}=\argmax_{j}\{z_j\}$.  
The constraint in~\autoref{constraints_1} means that $z_l + e_l$ is the largest entry in the vector $\mathbf{z+e}$. Therefore, we enforce the inequality constraint
$z_l + e_l \geq max_{j|j\neq l}\{z_j+e_j\}$. Moreover, we further transform the inequality constraint to the following loss function:
\begin{equation}
L_2=\text{ReLU}(-z_l-e_l+max_{j|j\neq l}\{z_j+e_j\}),
\end{equation}
where the function ReLU is defined as ReLU$(v)$=$\max\{0, v\}$. 
The loss function $L_2$ is 0 if the inequality $z_l + e_l \geq max_{j|j\neq l}\{z_j+e_j\}$ holds. 

\myparatight{Unconstrained optimization problem} After transforming the constraints into the objective function, we have the following unconstrained optimization problem:
\begin{align}
\min_{\mathbf{e}}\  L=L_1+c_2\cdot L_2 + c_3\cdot L_3,
\end{align}
where $L_3=d(softmax(\mathbf{z}),softmax(\mathbf{z}+\mathbf{e}))$, while $c_2$ and $c_3$ balance between the three terms.

\myparatight{Solving the unconstrained optimization problem} We design an algorithm based on gradient descent to solve the unconstrained optimization problem. Algorithm~\autoref{algorithml1} shows our algorithm. Since we aim to find a noise vector that has a small confidence score distortion, we iteratively search a large $c_3$. For each given $c_3$, we use gradient descent to find $\mathbf{e}$ that satisfies the constraints in \autoref{constraints_1} and~\autoref{constraints_2}. The process of searching $c_3$ stops when we cannot find a vector  $\mathbf{e}$ that satisfies the two constraints. Specifically, given $c_2$, $c_3$, and a learning rate $\beta$, we iteratively update the vector variable $\mathbf{e}$ (i.e., the inner while loop in Algorithm~\autoref{algorithml1}). Since we transform the constraints in \autoref{constraints_1} and~\autoref{constraints_2} into the objective function, there is no guarantee that they are satisfied during the iterative gradient descent process. Therefore, in each iteration of gradient descent, we check whether the two constraints are satisfied (i.e., Line~\autoref{conditioncheck} in Algorithm~\autoref{algorithml1}). Specifically, we continue the gradient descent process when the predicted label changes or the sign of the logit $h$ does not change. In other words, we stop the gradient descent process when both constraints are satisfied. We use $h(softmax(\mathbf{z}))\cdot h(softmax(\mathbf{z}+\mathbf{e})) \leq 0$ to approximate the constraint in~\autoref{constraints_2}. In particular, the constraint in~\autoref{constraints_2} is equivalent to $h(softmax(\mathbf{z}+\mathbf{e}))=0$.   Once we find a vector $\mathbf{e}$ such that $h(softmax(\mathbf{z}))$ and $h(softmax(\mathbf{z}+\mathbf{e}))$ have different signs (e.g., $h(softmax(\mathbf{z})) >0$ and $h(softmax(\mathbf{z}+\mathbf{e}))<0$), $h(softmax(\mathbf{z}+\mathbf{e}))$ just crosses 0 and should be close to 0 since we use a small learning rate. Note that we could also iteratively search $c_2$, but it is computationally inefficient to search both $c_2$ and $c_3$.

\subsection{Phase II}
After Phase I, we have two representative noise vectors. One is $\mathbf{0}$ and the other  is $\mathbf{r}$. In Phase II, we assume the randomized noise addition mechanism is a probability distribution over the two representative noise vectors instead of the entire noise space. Specifically, we assume that the defender picks the representative noise vectors $\mathbf{r}$ and $\mathbf{0}$ with probabilities $p$ and $1-p$, respectively; and the defender adds the picked representative noise vector to the true confidence score vector.  With such simplification, we can simplify the optimization problem in~\autoref{originalproblem} to the following optimization problem:
\begin{align}
    p=\argmin_{p}& |p\cdot g(\mathbf{s}+\mathbf{r}) + (1-p)\cdot g(\mathbf{s+0})-0.5| \\
    \text{subject to: }&p\cdot d(\mathbf{s},\mathbf{s}+\mathbf{r}) + (1-p)\cdot d(\mathbf{s},\mathbf{s}+\mathbf{0}) \leq \epsilon,
\end{align}
where the constraint means that the expected confidence score distortion is bounded by the budget. Note that we omit the other three constraints in~\autoref{originalproblem-c1},~\autoref{originalproblem-c3}, and~\autoref{originalproblem-c4}. This is because both of our representative noise vectors already satisfy those constraints. Moreover, we can derive an analytical solution to the simplified optimization problem. The analytical solution is as follows:
\begin{align}
\label{computep}
    p=\begin{cases}
     0, &\text{ if }|g(\mathbf{s})-0.5|\leq |g(\mathbf{s}+\mathbf{r})-0.5|\\
    \min(\frac{\epsilon}{d(\mathbf{s},\mathbf{s}+\mathbf{r})},1.0), &\text{ otherwise.}
    \end{cases}
\end{align}

\myparatight{One-time randomness} If the defender randomly samples one of the two representative noise vectors every time for the same query data sample, then  an attacker could infer the true confidence score vector via querying the same data sample multiple times. We consider the attacker knows our defense mechanism including the confidence score distortion metric $d$, the budget $\epsilon$, and that the noise vector is sampled from two representative noise vectors, one of which is $\mathbf{0}$.
  
Suppose the attacker queries the same data sample $n$ times from the target classifier. The attacker receives a confidence score vector $\mathbf{s}_1$ for $m$ times and a confidence score vector $\mathbf{s}_2$ for $n-m$ times. One confidence score vector is $\mathbf{s+r}$ and the other is the true confidence score vector $\mathbf{s}$. Since the attacker receives two different confidence score vectors, the attacker knows $0<p<1$. Moreover, given the two confidence score vectors, the attacker can compute $p$ according to~\autoref{computep} since the distance $d(\mathbf{s},\mathbf{s}+\mathbf{r})$ does not depend on the ordering of $\mathbf{s}$ and $\mathbf{s}+\mathbf{r}$, i.e., $d(\mathbf{s},\mathbf{s}+\mathbf{r})=d(\mathbf{s}_1,\mathbf{s}_2)$. The attacker can also estimate the probabilities that the defender returns the confidence score vectors $\mathbf{s}_1$  and $\mathbf{s}_2$ as $\frac{m}{n}$ and $\frac{n-m}{n}$, respectively. If $\frac{m}{n}$ is closer to $p$, then the attacker predicts that $\mathbf{s}_2$ is the true confidence score vector, otherwise the attacker predicts $\mathbf{s}_1$ to be the true confidence score vector.

To address this challenge, we propose to use one-time randomness when the defender samples the representative noise, with which the defender always returns the same confidence score vector for the same query data sample. Specifically, 
for a query data sample, the defender quantizes each dimension of the query data sample and computes the hash value of the quantized data sample. Then, the defender generates a random number  $p^{\prime}$ in the range $[0,1]$ via a pseudo random number generator with the hash value as the seed. If $p^{\prime}<p$, the defender adds the representative noise vector $\mathbf{r}$ to the true confidence score vector, otherwise the defender does not add noise. The random number $p^{\prime}$ is the same for the same query data sample, so the defender always returns the same confidence score vector for the same query data sample. We compute the hash value of the quantized query data sample as the seed such that the attacker cannot just slightly modify the query data sample to generate a different $p^{\prime}$. 
The attacker can compute the random number $p^{\prime}$ as we assume the attacker knows the defense mechanism including the hash function and pseudo random number generator. However, the attacker does not know $p$ any more because the defender always returns the same confidence score vector for the same query data sample. Therefore, the attacker does not know whether the returned confidence score vector is the true one or not.

%% file: exp.tex
\section{Evaluation}
\label{evaluation_section}

\subsection{Experimental Setup}
\subsubsection{Datasets} We use three datasets that represent different application scenarios.

\myparatight{Location}  This dataset was preprocessed from the Foursquare dataset\footnote{https://sites.google.com/site/yangdingqi/home/foursquare-dataset} and we obtained  it from~\cite{SSSS17}. 
The dataset has 5,010 data samples with 446 binary features, each of which represents whether a user visited a particular region or location type.
 The data samples are grouped into $30$ clusters. This dataset represents a 30-class classification problem, where each cluster is a class.

\myparatight{Texas100} This dataset  is based on the Discharge Data public use files published by the Texas
Department of State Health Services.\footnote{https://www.dshs.texas.gov/THCIC/Hospitals/Download.shtm} We obtained the preprocessed dataset from~\cite{SSSS17}. 
The dataset has $67,330$ data samples with $6,170$ binary features. 
These features represent the external causes of injury (e.g., suicide, drug misuse), the diagnosis, the procedures the patient underwent, and some generic information (e.g., gender, age, and race). 
Similar to~\cite{SSSS17}, we focus on the $100$ most frequent procedures and the classification task is to predict a procedure for a patient using the patient's data. This dataset represents a 100-class classification problem.

\myparatight{CH-MNIST} This dataset is used for  classification of different tissue types on histology tile from  patients with colorectal cancer. The dataset contains $5,000$ images from $8$ tissues. The classification task is to predict tissue for an image, i.e., the dataset is a 8-class classification problem. The size of each image is $64\times 64$. 
We obtained a preprocessed version from Kaggle.~\footnote{\url{https://www.kaggle.com/kmader/colorectal-histology-mnist}}.

\myparatight{Dataset splits} 
For each dataset, we will train a target classifier, an attack classifier, and a defense classifier. Therefore, we split each dataset into multiple folds. Specifically, for the Location (or CH-MNIST) dataset, we randomly sample 4 disjoint sets, each of which includes 1,000 data samples. We denote them as ${D_1}$, ${D_2}$, ${D_3}$, and ${D_4}$, respectively.  For the  Texas100 dataset, we also randomly sample such 4 disjoint sets, but each set includes 10,000 data samples as the Texas100 dataset is around one order of magnitude larger.  Roughly speaking, for each dataset, we use ${D_1}$, ${D_2}$, and ${D_3}$ to learn the target classifier, the attack classifier, and the defense classifier, respectively; and we use  $D_1\cup{D_4}$ to evaluate the accuracy of the attack classifier.  We will describe more details on how the sets are used when we use them.

\begin{table}[t]\renewcommand{\arraystretch}{1}
    \centering
    \caption{Neural network architecture of the target classifier for CH-MNIST.}
    \begin{tabular}{|c|c|} \hline 
    Layer Type & Layer Parameters  \\ \hline
    \multicolumn{2}{|c|}{Input  $64\times 64$} \\ \hline
    Convolution& $32\times 3 \times 3$, strides=$(1, 1)$, padding=same  \\ 
    Activation& ReLU  \\ \hline
    Convolution& $32\times 3 \times 3$, strides=$(1, 1)$  \\ 
    Activation& ReLU  \\ 
    Pooling& MaxPooling$(2\times 2)$ \\ \hline
    Convolution& $32\times 3 \times 3$, strides=$(1, 1)$, padding=same  \\ 
    Activation& ReLU  \\ \hline
    Convolution& $32\times 3 \times 3$, strides=$(1, 1)$  \\ 
    Activation& ReLU  \\ \hline
    Pooling& MaxPooling$(2\times 2)$  \\ \hline
    Flatten&   \\ \hline
    Fully Connected& 512  \\ \hline
    Fully Connected& 8  \\ \hline
    Activation& softmax  \\ \hline
    \multicolumn{2}{|c|}{Output} \\ \hline
    \end{tabular} 
    \label{architecture_ch_mnist} 
\end{table}

\subsubsection{Target Classifiers} For the Location and Texas100 datasets, we use a fully-connected neural network with $4$ hidden layers as the target classifier. The number of neurons for the four layers are 1024, 512, 256, and 128, respectively. We use the popular activation function ReLU for the neurons in the hidden layers. The activation function in the output layer is softmax. We adopt the cross-entropy loss function and use Stochastic Gradient Descent (SGD) to learn the model parameters. We train $200$ epochs with a learning rate $0.01$, and we decay the learning rate by $0.1$ in the $150$th epoch for better convergence. For the CH-MNIST dataset, the neural network architecture of the target classifier is shown in~\autoref{architecture_ch_mnist}. Similarly, we also adopt the cross-entropy loss function and use SGD to learn the model parameters. We train $400$ epochs with a learning rate $0.01$ and decay the learning rate by $0.1$ in the $350$th epoch. 
For each dataset, we use $D_1$ to train the target classifier.~\autoref{accuracy_of_target_classifier} shows the training and testing accuracies of the target classifiers on the three datasets, where the testing accuracy is calculated by using the target classifier to make predictions for the data samples that are not in $D_1$.

\subsubsection{Membership Inference Attacks} 
\label{membershipattack}
In a membership inference attack, an attacker trains an attack classifier, which predicts \emph{member} or \emph{non-member} for a query data sample. The effectiveness of an attack is measured by the \emph{inference accuracy} of the attack classifier, where the inference accuracy is the fraction of data samples in $D_1\cup{D_4}$ that the attack classifier can correctly predict as member or non-member. In particular, data samples in $D_1$ are members of the target classifier's training dataset, while data samples in $D_4$ are non-members. We call the dataset $D_1\cup{D_4}$ \emph{evaluation dataset}. We consider two categories of state-of-the-art black-box membership inference attacks, i.e., \emph{non-adaptive attacks} and \emph{adaptive attacks}. In non-adaptive attacks, the attacker does not adapt its attack classifier based on our defense, while the attacker adapts its attack classifier based on our defense in adaptive attacks.


\begin{figure*}[!t]
	 \centering
	 \subfloat[Location]{\includegraphics[width=0.330\textwidth]{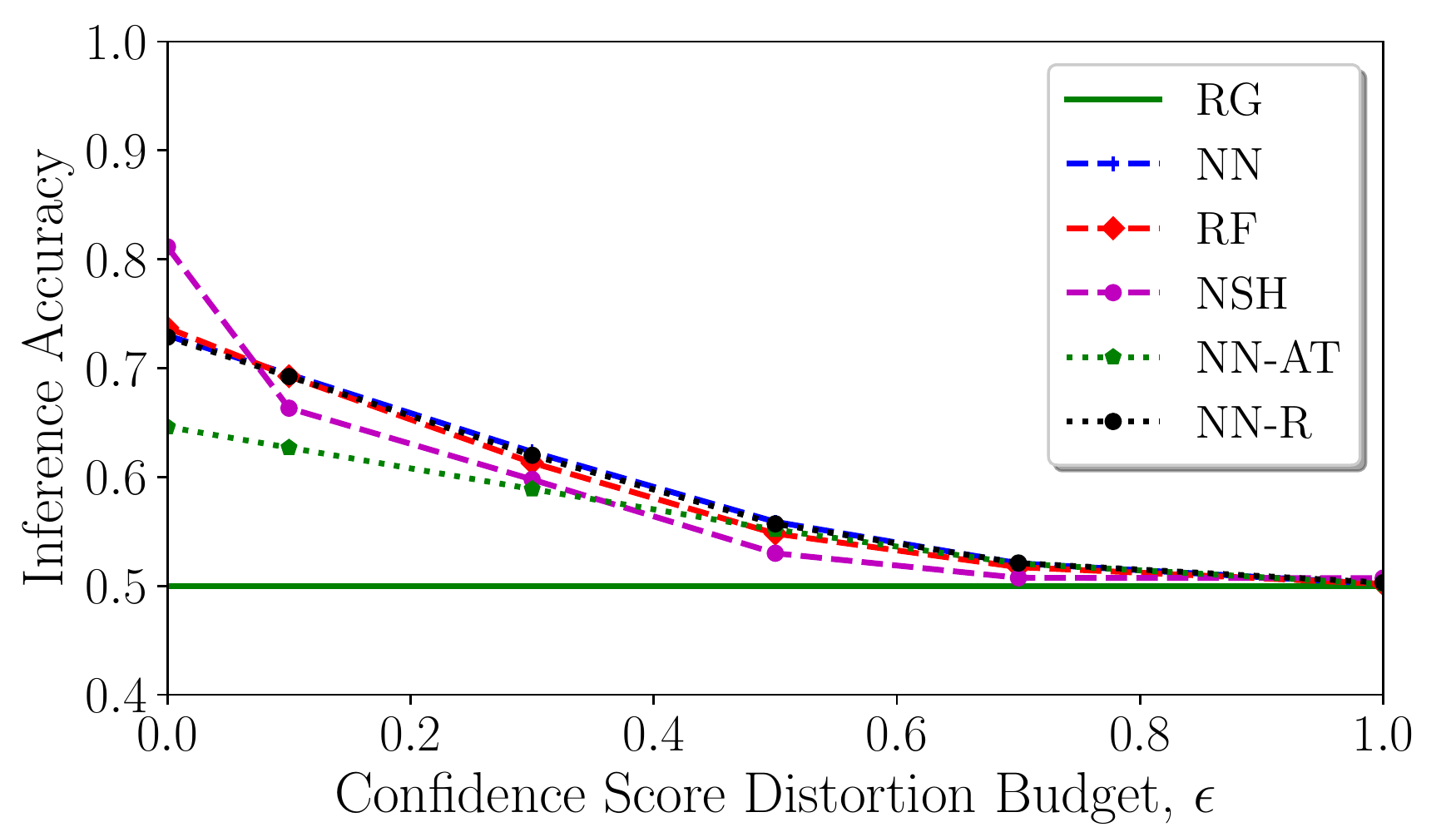}}
	 \subfloat[Texas100]{\includegraphics[width=0.330\textwidth]{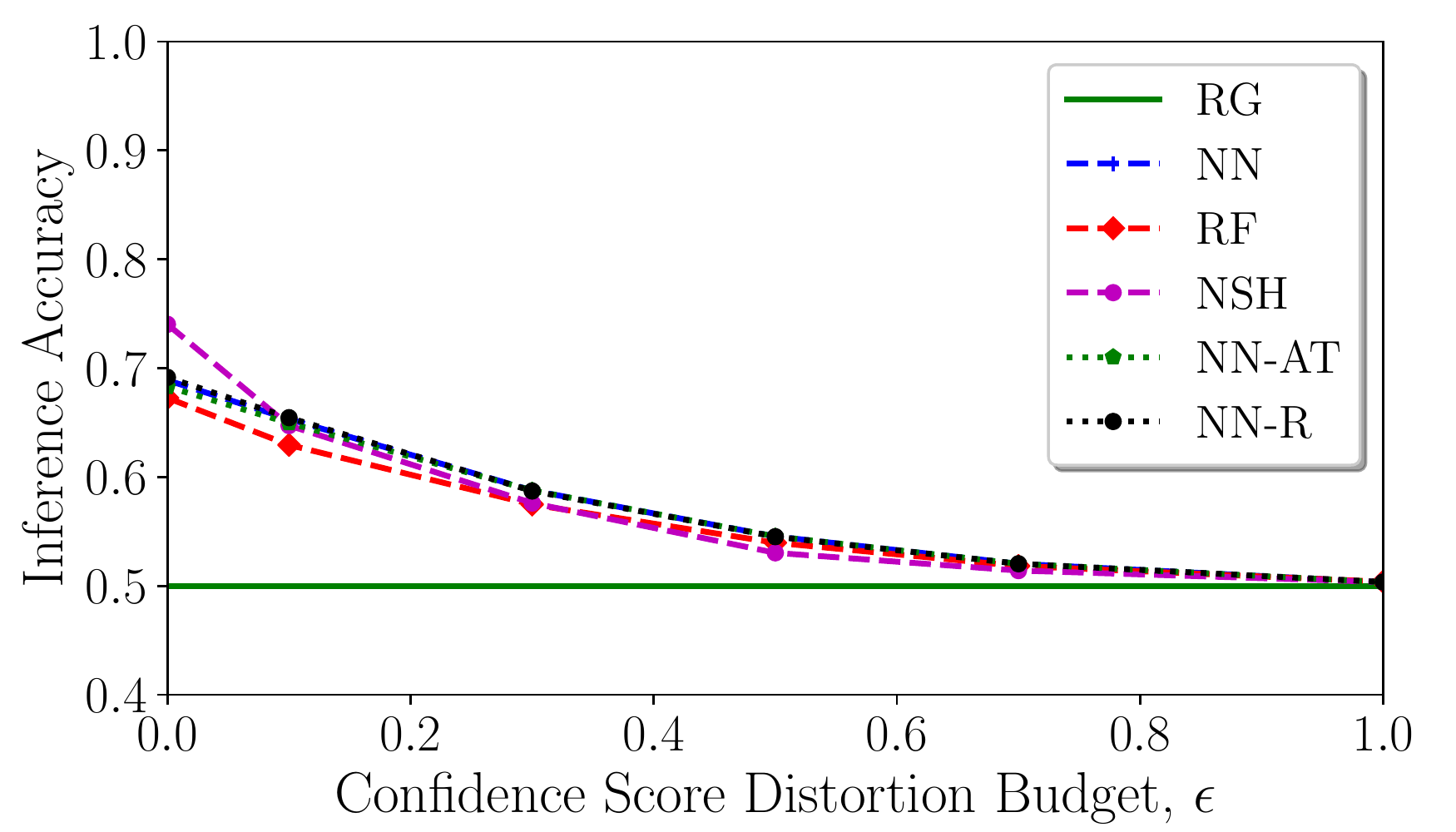}}
	 \subfloat[CH-MNIST]{\includegraphics[width=0.330\textwidth]{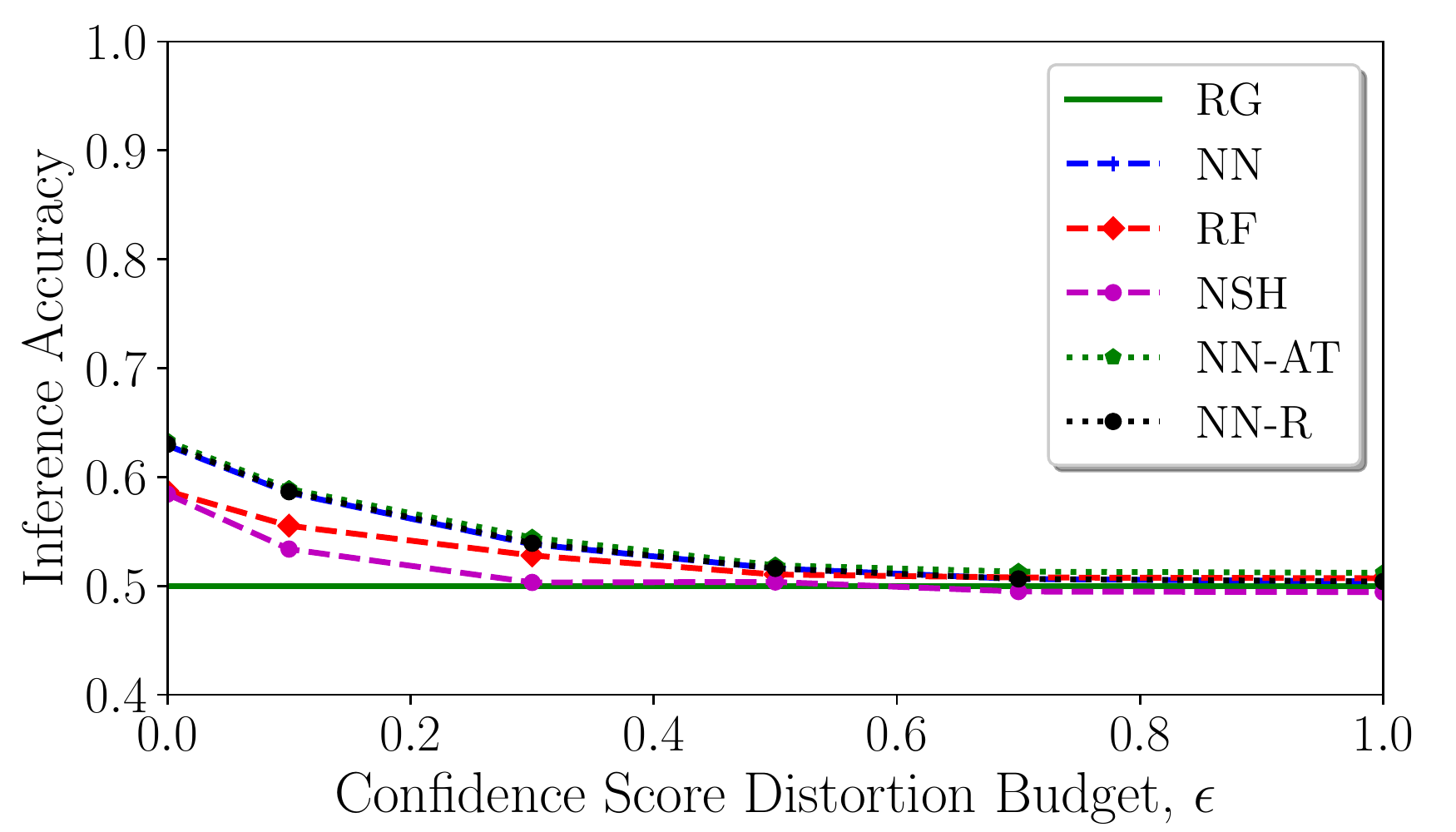}}	 
	 \caption{Inference accuracies of different attacks as the confidence score distortion budget (i.e., $\epsilon$) increases.}
	 \label{fix_k_query_infer_acc}
\end{figure*}

\myparatight{Non-adaptive attacks} We consider the \emph{random guessing} attack and state-of-the-art attacks as follows. 

 {\bf Random guessing (RG) attack}.  For any query data sample, this attack predicts it to be a member of the target classifier's training dataset with probability 0.5. The inference accuracy of the RG attack is 0.5. 

 \begin{table}
	\centering
	\caption{Training and testing accuracies of the target classifier on the three datasets.}
	\begin{tabular}{|c|c|c|c|} 
	\hline
			 & Location & Texas100 & CH-MNIST  \\ 
	\hline
	Training Accuracy &   100.0\%      &   99.98\%     &    99.0\%      \\
	\hline
	Testing Accuracy &   60.32\%      &     51.59\%    &   72.0\%      \\ 
	\hline
	\end{tabular}
	\label{accuracy_of_target_classifier}
\end{table}

 {\bf Neural Network (NN) attack~\cite{SSSS17,SZHBFB19}}. This attack assumes that the attacker knows the distribution of the target classifier's training dataset and the architecture of the target classifier.  We further split the dataset $D_2$ into two halves denoted as $D_2'$ and $D_2''$, respectively. The attacker uses $D_2'$ to train a shadow classifier that has the same neural network architecture as the target classifier. After training the shadow classifier, the attacker calculates the confidence score vectors for the data samples in $D_2'$ and $D_2''$, which are members and non-members of the shadow classifier. Then, the attacker ranks each confidence score vector and treats the ranked confidence score vectors of members and non-members as a ``training dataset'' to train an attack classifier. The attack classifier takes a data sample's ranked confidence score vector as an input and predicts member or non-member.  For all three datasets, we consider the attack classifier is a fully-connected neural network with three hidden layers, which have 512, 256, and 128 neurons, respectively. The output layer just has one neuron. The neurons in the hidden layers use the ReLU activation function, while the neuron in the output layer uses the sigmoid activation function. The attack classifier predicts member if and only if the neuron in the output layer outputs a value that is larger than 0.5. We train the attack classifier for 400 epochs with a learning rate 0.01 using SGD and decay the learning rate by 0.1 at the 300th epoch.

 {\bf Random Forest (RF) attack}. This attack is the same as the NN attack except that RF attack uses random forest as the attack classifier, while NN uses a neural network as the attack classifier.  We use scikit-learn with the default setting to learn random forest classifiers. We consider this RF attack to demonstrate that our defense mechanism is still effective even if the attack classifier and the defense classifier (a neural network) use different types of algorithms, i.e., the noise vector that evades the defense classifier can also evade the attack classifier even if the two classifiers use different types of algorithms.

 {\bf NSH attack~\cite{NSH18}}. Nasr, Shokri, and Houmansadr~\cite{NSH18} proposed this attack, which we abbreviate as NSH. This attack uses multiple neural networks. One network operates on the confidence score vector. Another one operates on the label which is one hot encoded. Both networks are fully-connected and have the same number of input dimension, i.e., the number of classes of the target classifier. Specifically, NSH assumes the attacker knows some members and non-members of the target classifier's training dataset. In our experiments, we assume the attacker knows $30\%$ of data samples in $D_1$ (i.e., members) and 30\% of data samples in $D_4$ (i.e., non-members). The attacker uses these data samples to train the attack classifier. We adopt the neural network architecture in~\cite{NSH18}  as the attack classifier. The remaining 70\% of data samples in $D_1$ and $D_4$ are used to calculate the inference accuracy of the attack classifier. We train the attack classifier for $400$ epochs with an initial learning rate $0.01$ and decay the learning rate by $0.1$ after  $300$ epochs. 
 

\myparatight{Adaptive attacks} We consider two attacks that are customized to our defense. 

{\bf Adversarial training (NN-AT)}.  One adaptive attack is to train the attack classifier via adversarial training, which was considered to be the most empirically robust method against adversarial examples so far~\cite{ACW18}.  We adapt the NN attack using adversarial training and denote the adapted attack as NN-AT. Specifically, for each data sample in $D_2'$ and $D_2''$, the attacker calculates its confidence score vector using the shadow classifier. Then, the attacker uses the Phase I of our defense to find the representative noise vector and adds it to the confidence score vector to obtain a noisy confidence score vector. Finally, the attacker trains the attack classifier via treating the true confidence score vectors and their corresponding noisy versions of data samples in $D_2'$ and $D_2''$ as a training dataset. 

{\bf Rounding (NN-R)}. Since our defense adds carefully crafted small noise to the confidence score vector, an adaptive attack is to \emph{round} each confidence score before using the attack classifier to predict member/non-member. Specifically, we consider the attacker rounds each confidence score to be one decimal and uses the NN attack. Note that rounding is also applied when training the NN attack classifier. We denote this attack NN-R.

\autoref{accuracy_without_defense} shows the inference accuracies of different attacks when our defense is not used. All attacks except RG have inference accuracies that are larger or substantially larger than 0.5.

\begin{table}
\centering
\caption{Inference accuracies of different attacks on the three datasets when our defense is not used.}
\begin{tabular}{|c|c|c|c|} 
\hline
         & Location & Texas100 & CH-MNIST  \\ 
\hline
RG &   50.0\%      &    50.0\%     &    50.0\%      \\
\hline
NN &   73.0\%     &      68.9\%   &      62.9\%    \\
\hline
RF &   73.7\%      &     67.3\%    &    58.7\%      \\ 
\hline
NSH &     81.1\%    &   74.0\%      &      58.4\%    \\
\hline
NN-AT &   64.6\%      &   68.3\%      &      63.3\%    \\ 
\hline
NN-R &   72.9\%      &   69.2\%      &      63.0\%    \\ 
\hline
\end{tabular}
\label{accuracy_without_defense}
\end{table}

\subsubsection{Defense Setting} \label{defensesetting}
In our defense, we need to specify a defense classifier and the parameters in Algorithm~\autoref{algorithml1}.

\myparatight{Defense classifier} The defender itself trains a classifier to perform membership inference. We consider the defense classifier is a neural network. However, since the defender does not know the attacker's attack classifier, we 
assume the defense classifier and the attack classifier use different neural network architectures. Specifically, we consider three different defense classifiers in order to study the impact of defense classifier on MemGuard. The three defense classifiers are fully-connected neural networks with 2, 3, and 4 hidden layers, respectively. The hidden layers of the three defense classifiers have (256, 128), (256, 128, 64), and (512, 256, 128, 64) neurons, respectively. 
The output layer has just one neuron. The activation function for the neurons in the hidden layers is $ReLU$, while the neuron in the output layer uses the sigmoid activation function. 
Unless otherwise mentioned, we use the defense classifier with 3 hidden layers.
 The defender calculates the confidence score vector for each data sample in $D_1$ and $D_3$ using the target classifier. The confidence score vectors for data samples in $D_1$ and $D_3$ have labels ``member'' and ``non-member'', respectively. The defender treats these confidence score vectors as a training dataset to learn a defense classifier, which takes a confidence score vector as an input and predicts member or non-member. We train a defense classifier for $400$ epochs with a learning rate $0.001$. We note that we can also synthesize data samples based on $D_1$ as non-members (Appendix~\ref{synthesizenonmembers} shows details).  

\myparatight{Parameter setting} We set $max\_iter=300$ and $\beta=0.1$ in Algorithm~\autoref{algorithml1}. We found that once $max\_iter$ is larger than some threshold, MemGuard's effectiveness does not change. Since we aim to find representative noise vector that does not change the predicted label, we assign a relatively large value to $c_2$, which means that the objective function has a large value if the predicted label changes (i.e., the loss function $L_2$ is non-zero). In particular, we set $c_2=10$. Our Algorithm~\autoref{algorithml1} searches for a large $c_3$ and we set the initial value of $c_3$ to be 0.1. We also compare searching $c_2$ with searching $c_3$.

\subsection{Experimental Results}
\label{experimentalresults}

\begin{figure*}[!t]
	 \centering
	 \subfloat[Location, without defense]{\includegraphics[width=0.330\textwidth]{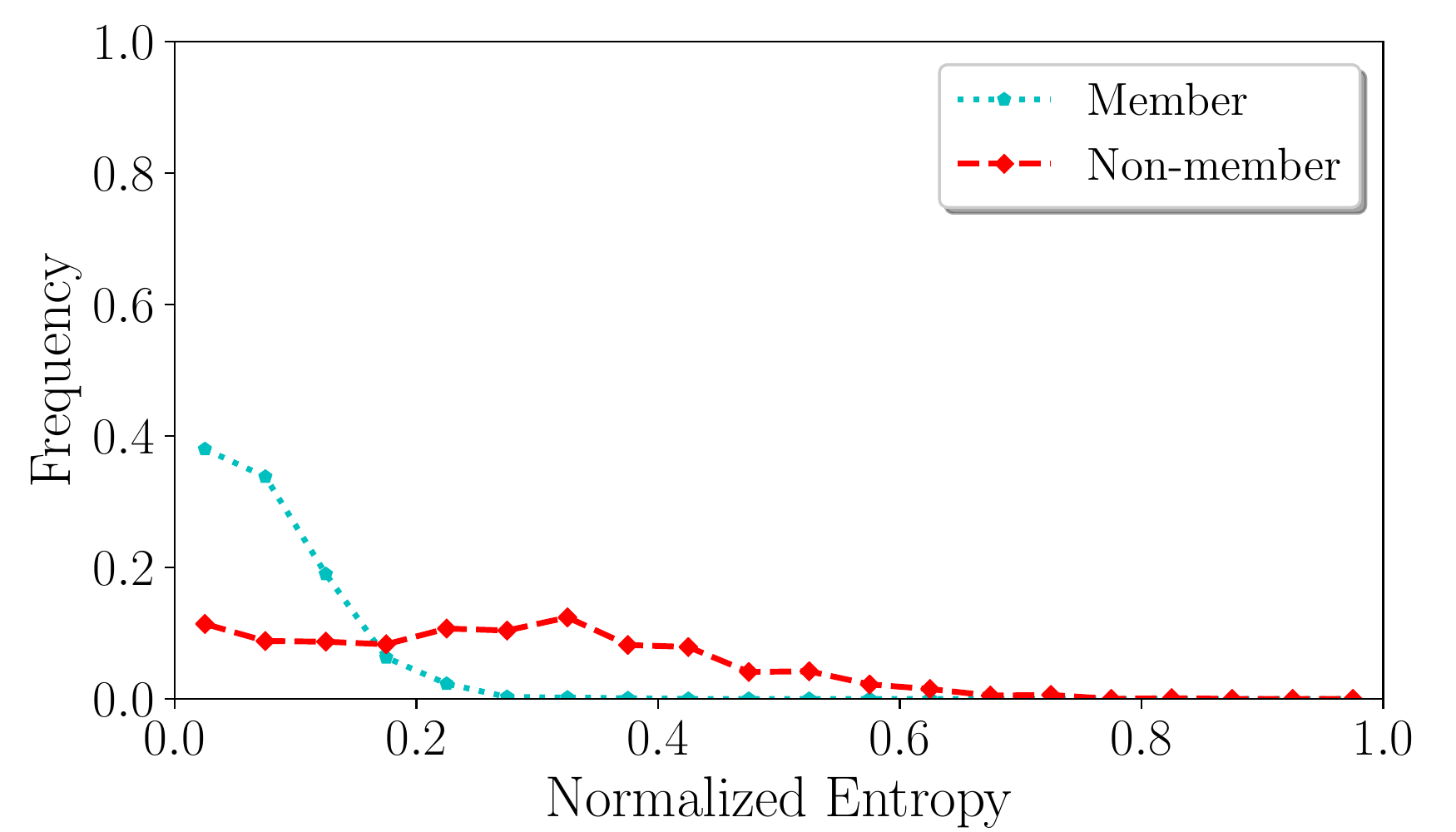}}
	 \subfloat[Texas100, without defense]{\includegraphics[width=0.330\textwidth]{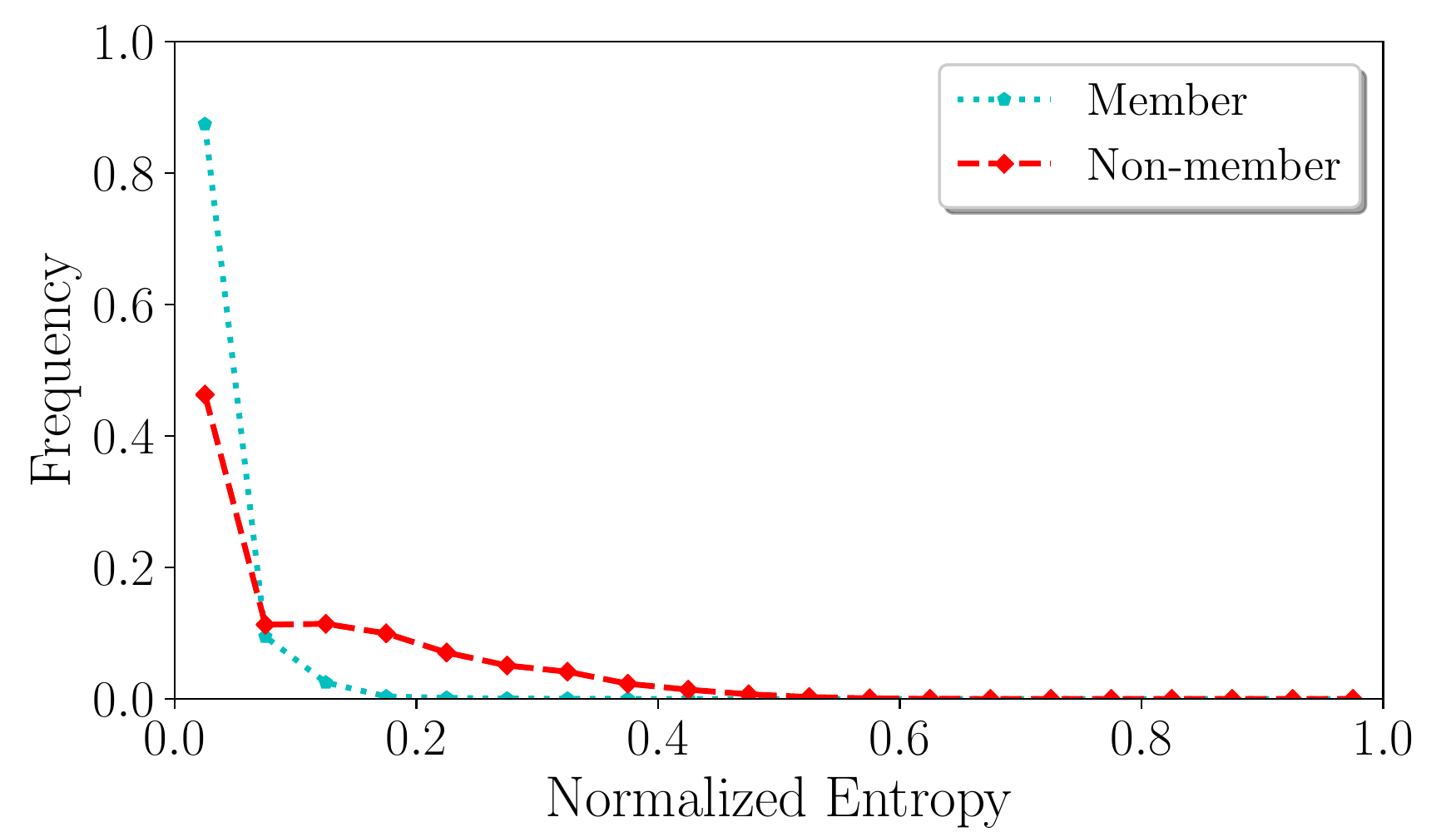}}
	 \subfloat[CH-MNIST, without defense]{\includegraphics[width=0.330\textwidth]{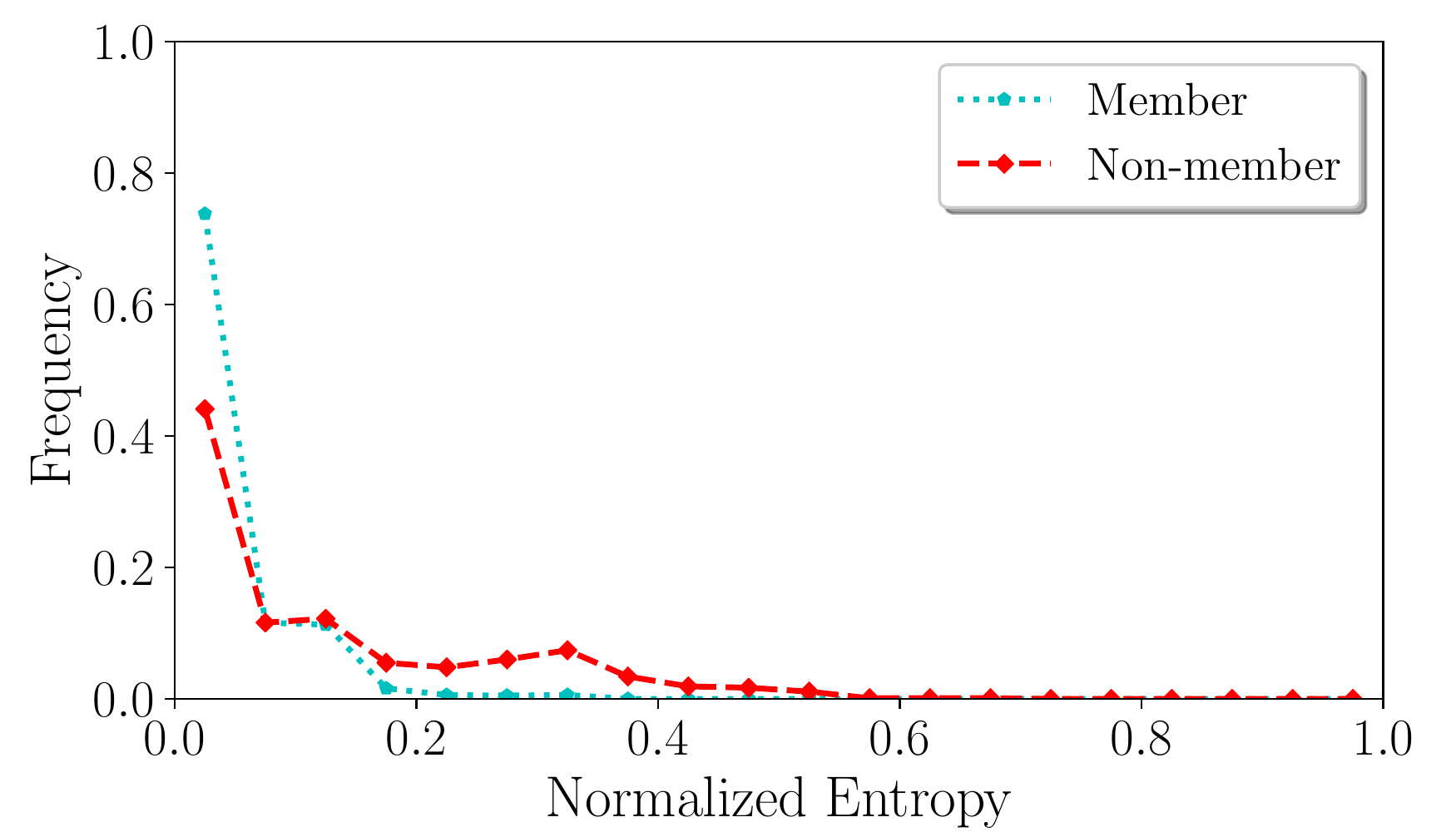}}

	 \subfloat[Location, with defense]{\includegraphics[width=0.330\textwidth]{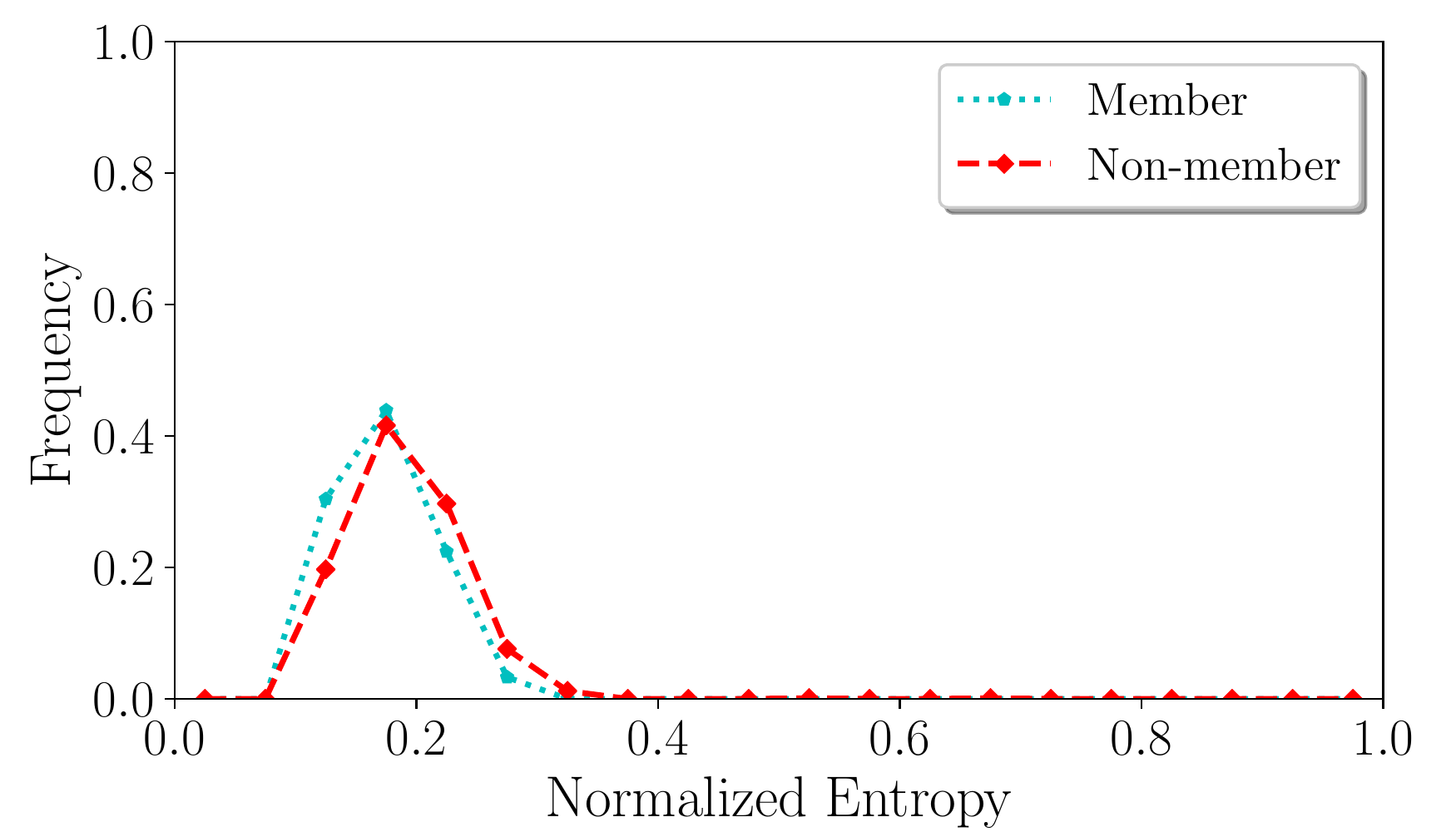}}
	 \subfloat[Texas100, with defense]{\includegraphics[width=0.330\textwidth]{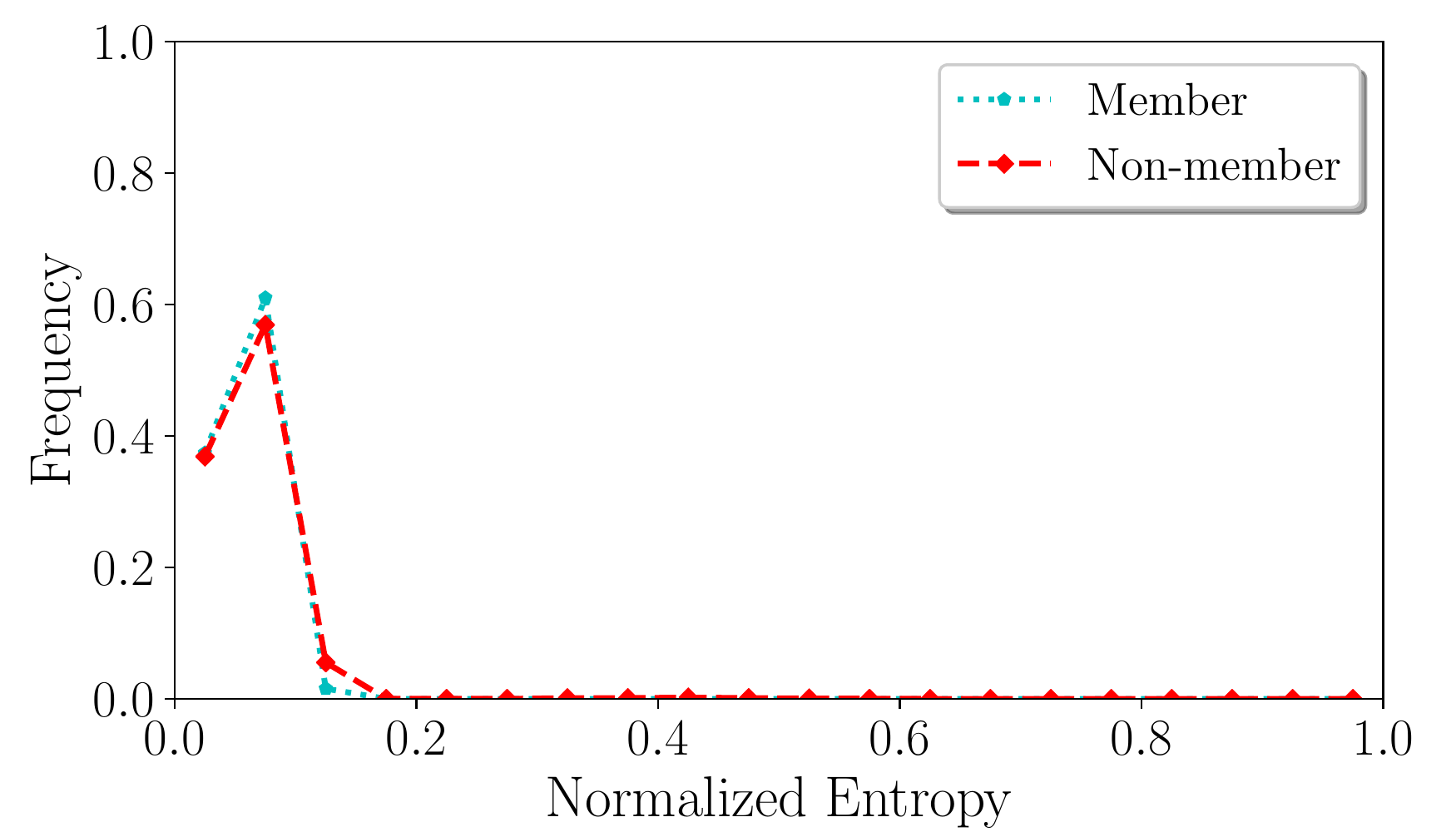}}	 	
	 \subfloat[CH-MNIST, with defense]{\includegraphics[width=0.330\textwidth]{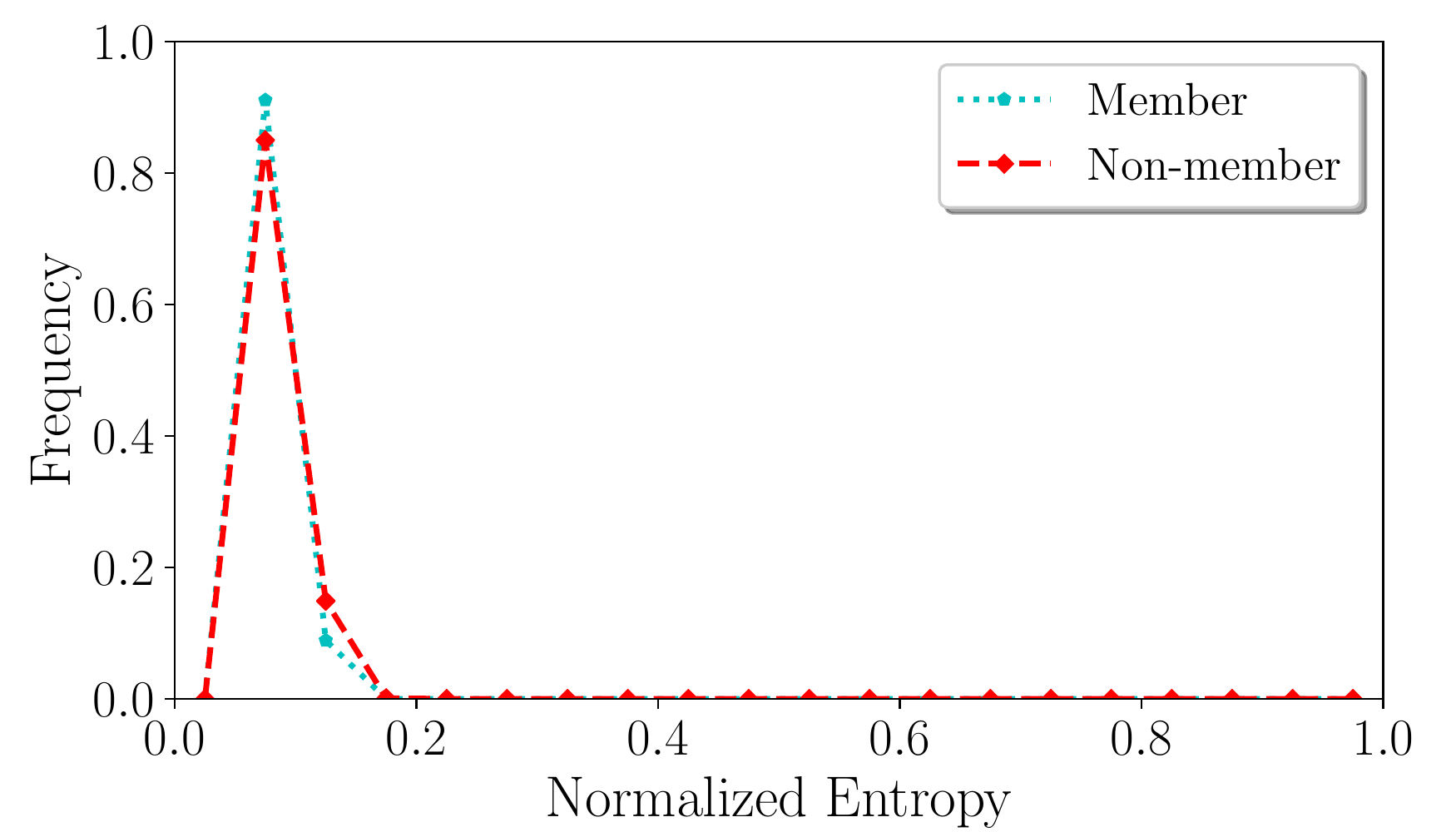}}
	 \caption{Distribution of the normalized entropy of the confidence score vectors for members and non-members of the target classifier. Figures on the upper side are results without our defense, and figures on the lower side are results with our defense.}
	 \label{entropy_distribution}
\end{figure*}

\myparatight{MemGuard is effective} ~\autoref{fix_k_query_infer_acc} shows the inference accuracies of different attacks as the confidence score distortion budget increases on the three datasets. Since we adopt the expected $L_1$-norm of the noise vector to measure the confidence score distortion, the confidence score distortion is in the range [0, 2]. Note that our defense is guaranteed to achieve 0 label loss as our Algorithm~\ref{algorithml1} guarantees that the predicted label does not change when searching for the representative noise vector. We observe that our MemGuard can effectively defend against membership inference attacks, i.e., the inference accuracies of all the evaluated attacks decrease as our defense is allowed to add larger noise to the confidence score vectors. For instance, on  Location, when our defense is allowed to add noise whose expected $L_1$-norm is around 0.8, our defense can reduce all the evaluated attacks to the random guessing (RG) attack; on  CH-MNIST, 
our defense can reduce the NSH attack (or the remaining attacks) to random guessing when allowed to add noise whose expected  $L_1$-norm is around 0.3 (or 0.7).

\myparatight{Indistinguishability between the confidence score vectors of members and non-members} We follow previous work~\cite{NSH18} to study the distribution of confidence score vectors of members vs. non-members of the target classifier. Specifically, given a confidence score vector $\mathbf{s}$, we compute its \emph{normalized entropy} as follows:
\begin{align}
\text{\bf Normalized entropy: } -\frac{1}{\log k}\sum_{j}s_j\log(s_j),
\end{align}
where $k$ is the number of classes in the target classifier.   
Figure~\ref{entropy_distribution} shows the distributions of the normalized entropy of the confidence score vectors for members (i.e., data samples in $D_1$) and non-members (i.e., data samples in $D_4$) of the target classifier, where we set the confidence score distortion budget $\epsilon$ to be 1 when our defense is used.  The gap between the two curves in a graph corresponds to the information leakage of the target classifier's training dataset. Our defense substantially reduces such gaps.  Specifically, the \emph{maximum gap} between the two curves (without defense vs. with defense) is (0.27 vs. 0.11), (0.41 vs. 0.05), and (0.30 vs. 0.06) on the Location, Texas100, and CH-MNIST datasets, respectively. Moreover, the \emph{average gap} between the two curves (without defense vs. with defense) is (0.062 vs. 0.011), (0.041 vs. 0.005), and (0.030 vs. 0.006) on the three datasets, respectively.

\begin{figure}[!t]
	 \centering
	 \subfloat[Searching $c_3$]{\includegraphics[width=0.24\textwidth]{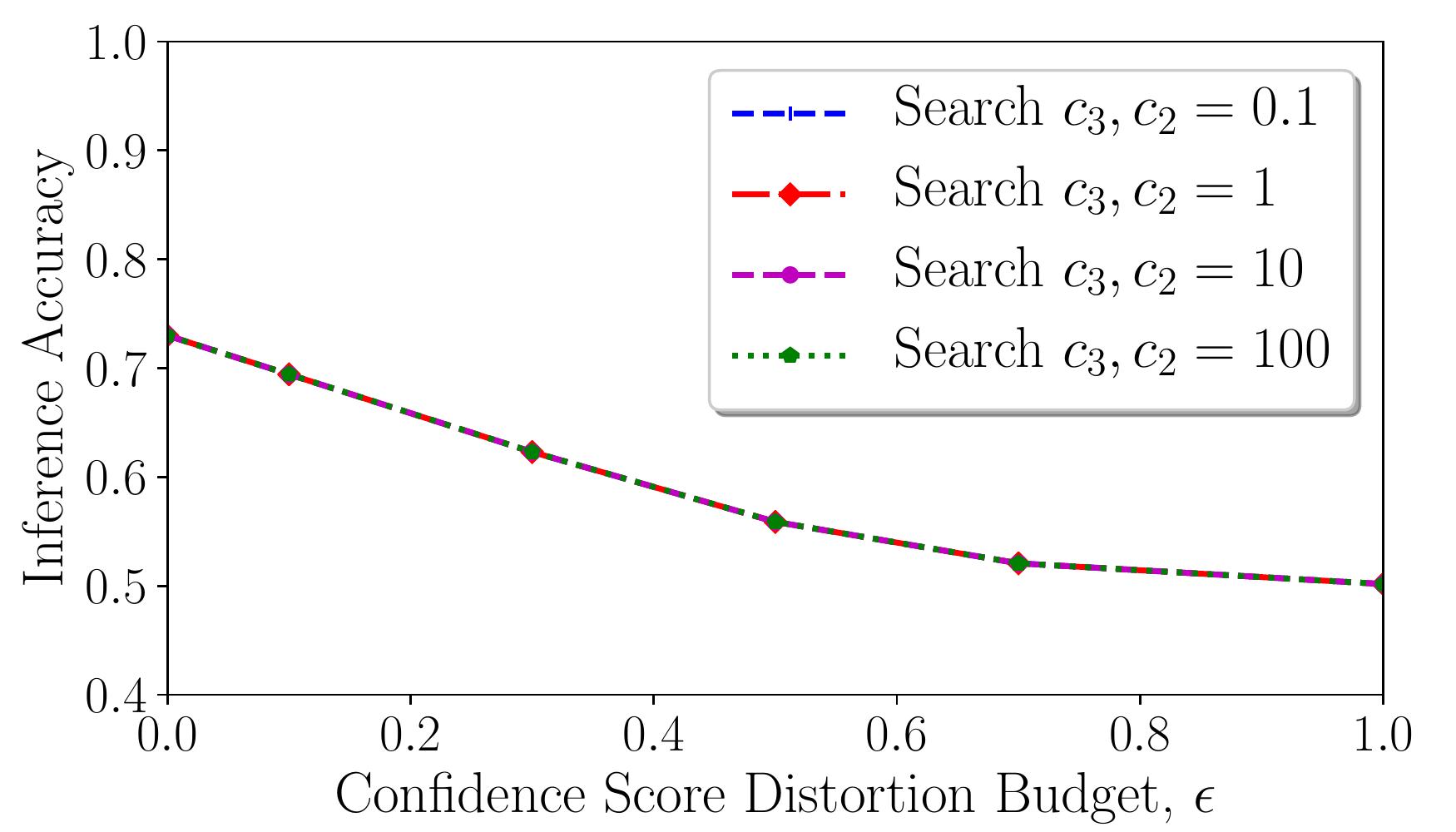}\label{searchc3-main}}
	 \subfloat[Searching $c_2$]{\includegraphics[width=0.24\textwidth]{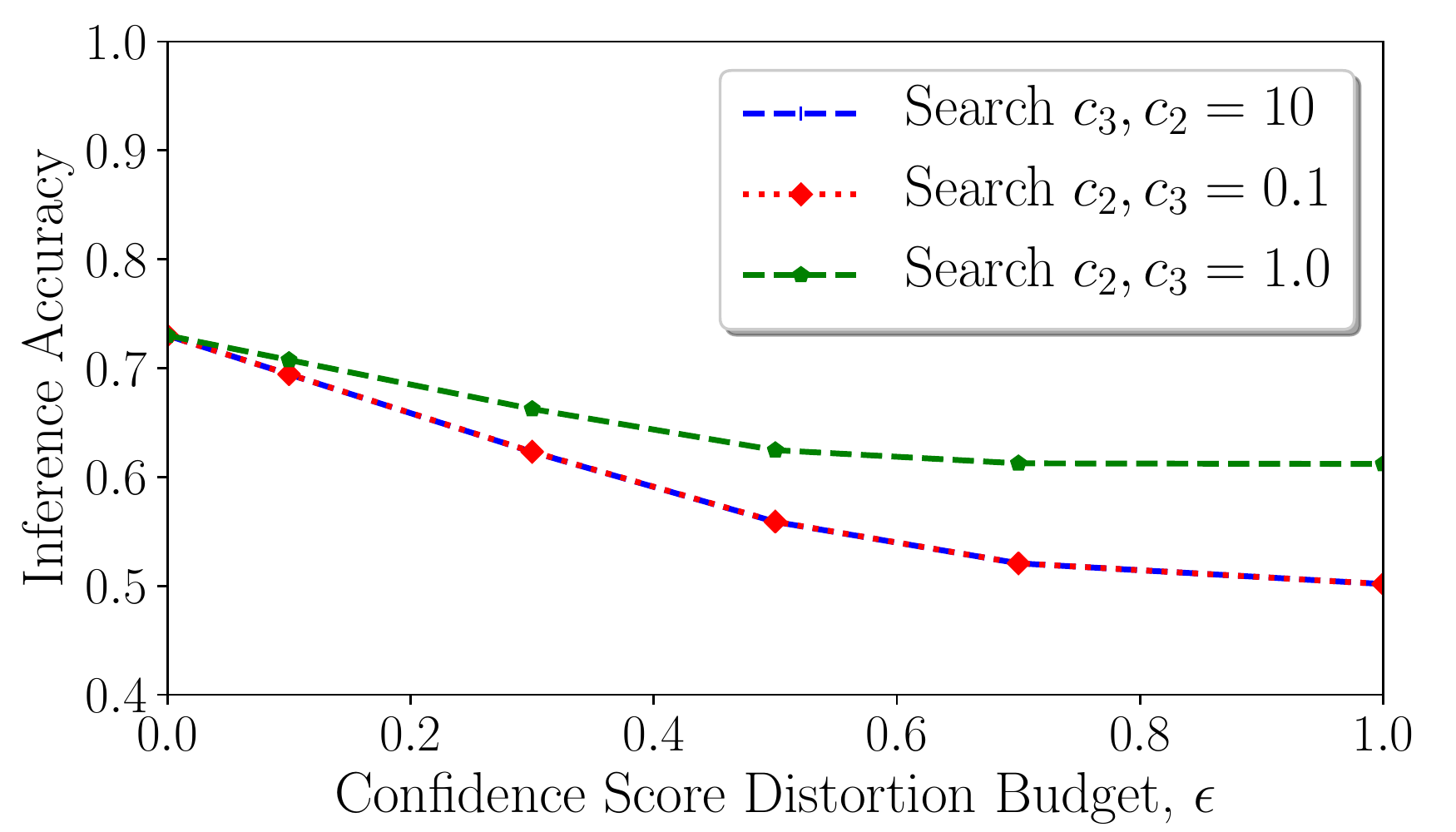}\label{searchc2-main}}
	 \caption{Inference accuracy of the NN attack as the confidence score distortion budget increases on the Location dataset when searching $c_3$ or $c_2$.}
\end{figure}

\myparatight{Searching $c_2$ vs. searching $c_3$} Figure~\ref{searchc3-main} shows the inference accuracy of the NN attack as the confidence score distortion budget  increases when fixing $c_2$ to different values and searching $c_3$. Figure~\ref{searchc2-main} shows the results when fixing $c_3$ and searching $c_2$. We observe that MemGuard is insensitive to the setting of $c_2$ when searching $c_3$. Specifically, MemGuard has almost the same effectiveness when fixing $c_2$ to different values, i.e., the different curves overlap in Figure~\ref{searchc3-main}. This is because when our Phase I stops searching the noise vector, the predicted label is preserved, which means that the loss function $L_2$ is 0.  However, MemGuard is sensitive to the setting of $c_3$ when searching $c_2$. Specifically, when fixing $c_3$ to be 0.1, searching $c_2$ achieves the same effectiveness as searching $c_3$. However, when fixing $c_3$ to be 1.0, searching $c_2$ is less effective. Therefore, we decided to search $c_3$ while fixing $c_2$.

\begin{figure}[!t]
	 \centering
	{\includegraphics[width=0.33\textwidth]{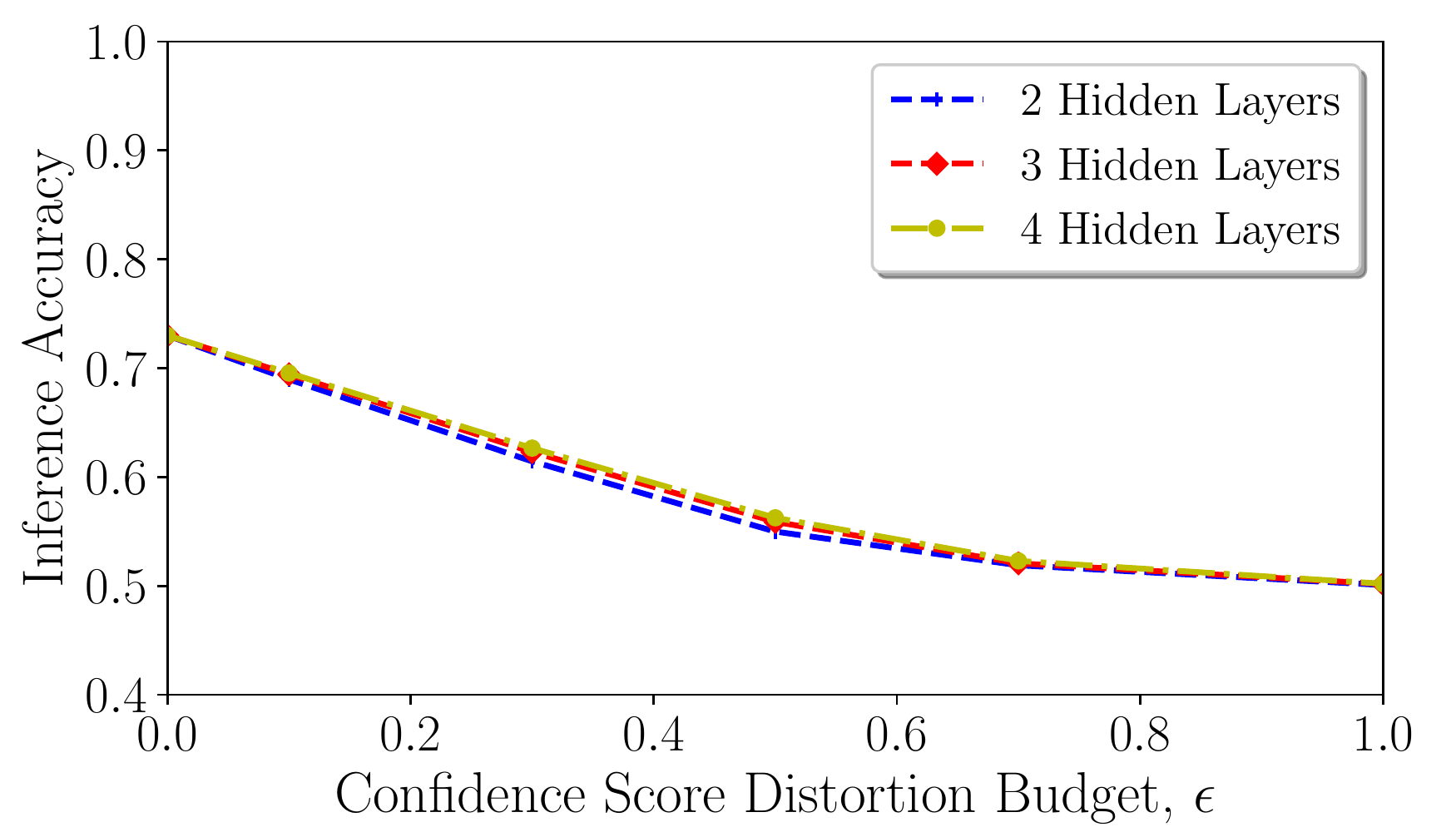}}
	 \caption{Inference accuracy of the NN attack as the confidence score distortion budget increases on the Location dataset when using different defense classifiers.}
	 \label{defenseclassifier}
\end{figure}

\myparatight{Impact of defense classifiers} Figure~\ref{defenseclassifier} shows the inference accuracy of the NN attack as the confidence score distortion budget increases on the Location dataset when using different defense classifiers. We observe that MemGuard has similar effectiveness for different defense classifiers, which means that our carefully crafted noise vectors can transfer between classifiers.

\myparatight{MemGuard outperforms existing defenses} We compare with state-of-the-art defenses including $L_2$-Regularizer~\cite{SSSS17}, Min-Max Game~\cite{NSH18}, Dropout~\cite{SZHBFB19}, Model Stacking~\cite{SZHBFB19}, and DP-SGD~\cite{ACGMMTZ16}. Each compared defense (except Model Stacking) has a hyperparameter to control the privacy-utility tradeoff. For instance, the hyperparameter that balances between the loss function and the $L_2$ regularizer in $L_2$-Regularizer, the hyperparameter that balances between the loss function and the adversarial regularizer in Min-Max Game, the dropout rate in Dropout, the privacy budget in DP-SGD, and $\epsilon$ in MemGuard. 
We also compare with MemGuard-Random in which we use the \emph{Random} method (refer to Section~\ref{sec:phaseI}) to generate the noise vector in Phase I. 

\begin{figure*}[!t]
	 \vspace{-2mm}
	 \centering
	 \subfloat[Location]{\includegraphics[width=0.330\textwidth]{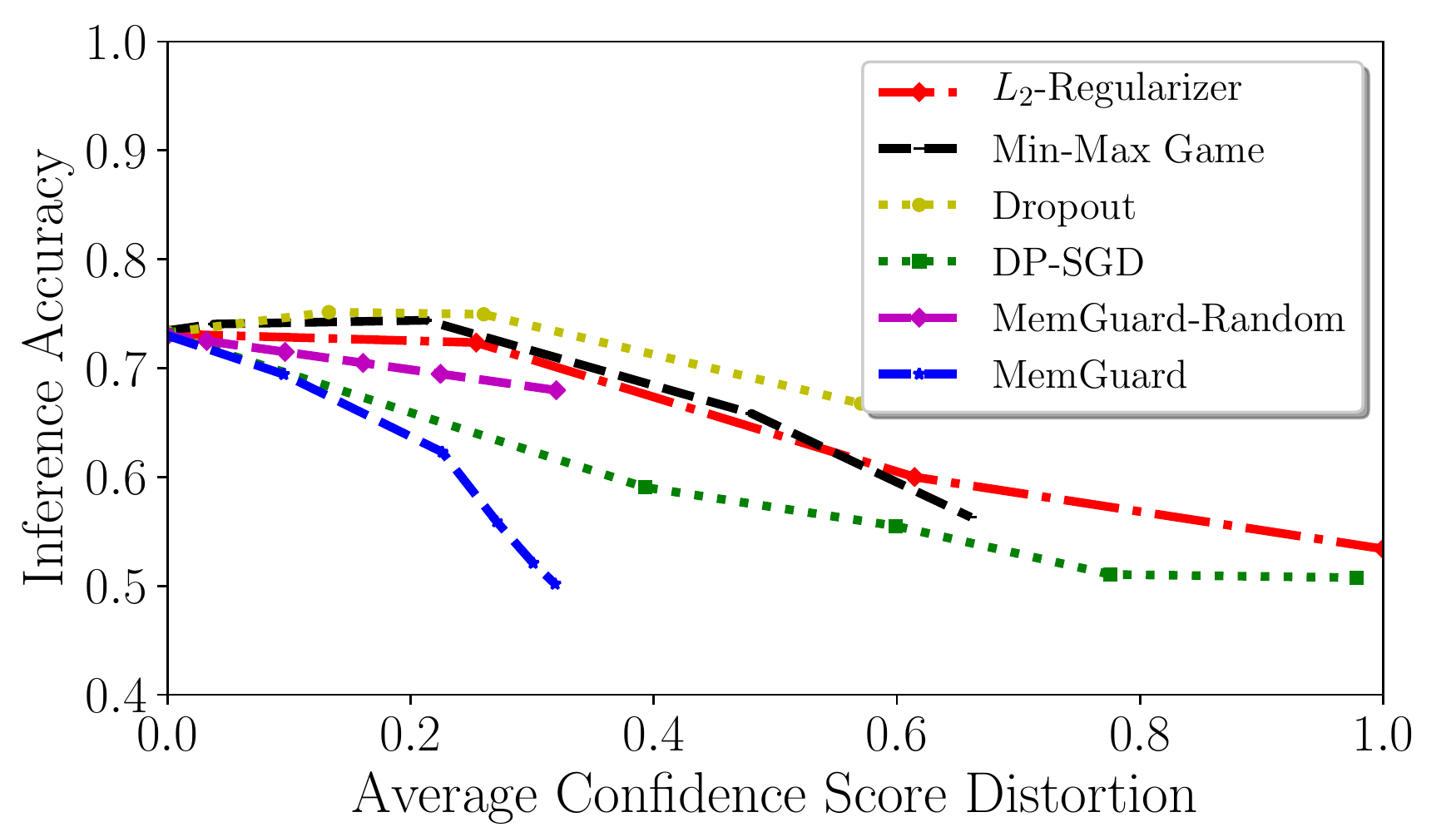}}
	 \subfloat[Texas100]{\includegraphics[width=0.330\textwidth]{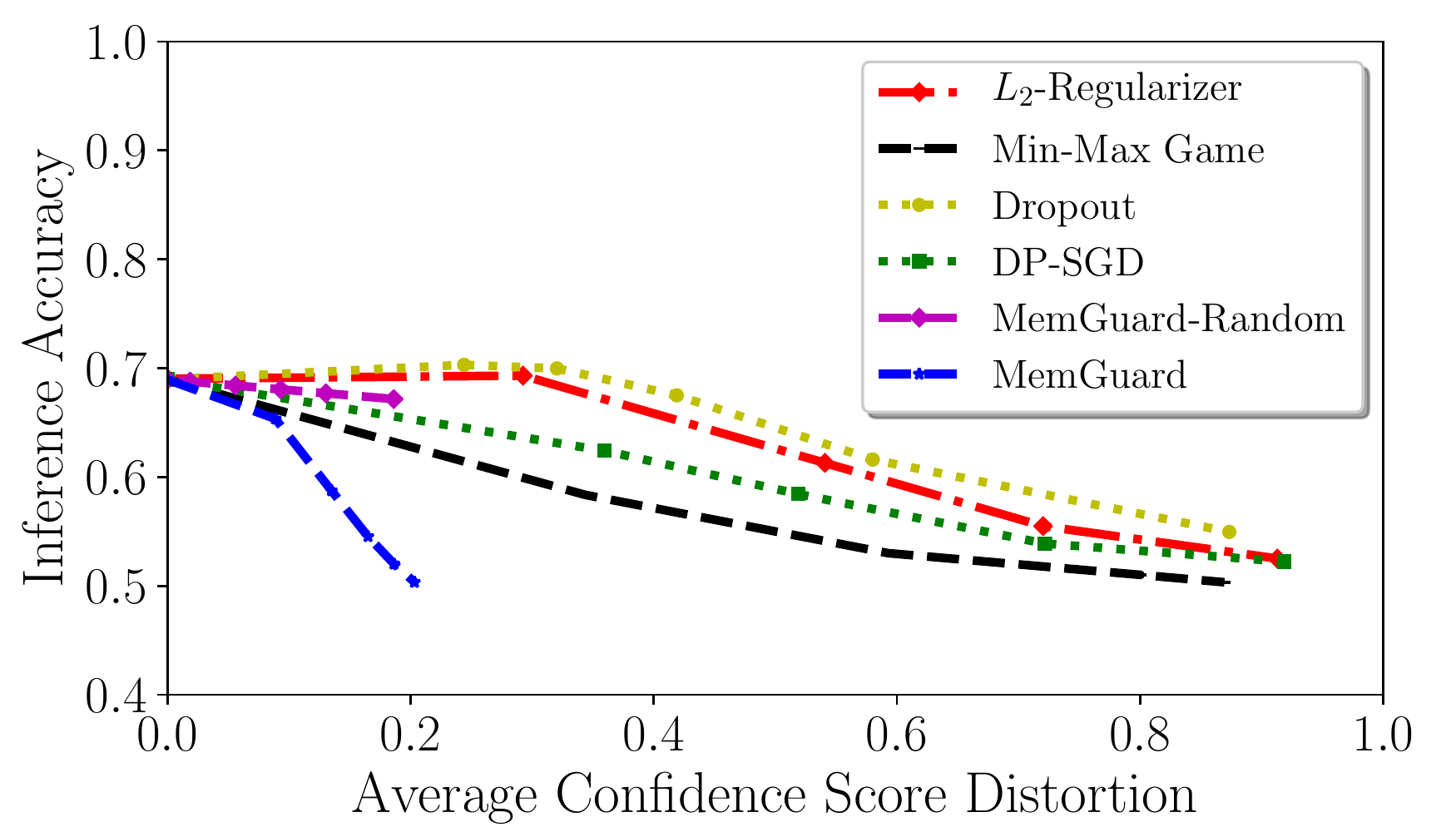}}
	 \subfloat[CH-MNIST]{\includegraphics[width=0.330\textwidth]{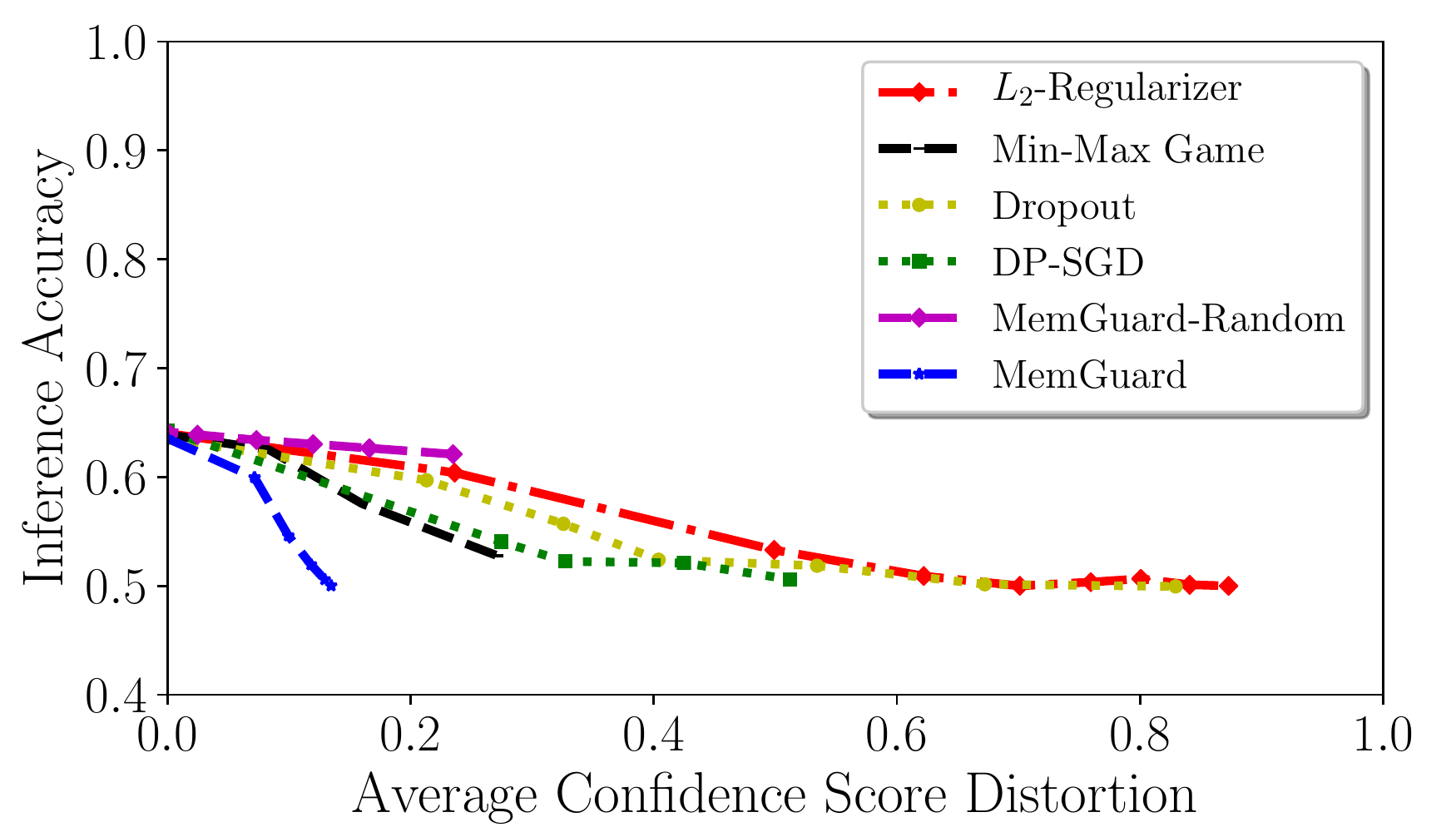}}
	 \caption{Inference accuracy vs. average confidence score distortion of the compared defenses. Our MemGuard achieves the best privacy-utility tradeoff.}
	 \label{fix_infer_acc}
\vspace{-1mm}
\end{figure*}

Before deploying any defense, we use the undefended target classifier to compute the confidence score vector for each data sample in the evaluation dataset $D_1\cup D_4$. For each defense and a given hyperparameter, we apply the defense to the target classifier and use the defended target classifier to compute the confidence score vector for each data sample in $D_1\cup D_4$. Then, we compute the confidence score distortion for each data sample and obtain the \emph{average confidence score distortion} on the evaluation dataset $D_1\cup D_4$. Moreover, we compute the inference accuracy of the attack classifier (we consider NN in these experiments) on the evaluation dataset after the defense is used. Therefore, for each defense and a given hyperparameter, we can obtain a pair (inference accuracy, average confidence score distortion). Via exploring different hyperparameters, we can obtain a set of such pairs for each defense. Then, we plot these pairs on a graph, which is shown in~\autoref{fix_infer_acc}.

Specifically, we tried the hyperparameter of $L_2$-Regularizer in the range $[0,0.05]$ with a step size $0.005$, $0.001$, and $0.005$ for Location, Texas100, and CH\_MNIST datasets, respectively.
We tried the hyperparameter of  Min-Max Game in the range $[0,3]$ with a step size $0.5$.
We tried the dropout rate of Dropout in the range $[0,0.9]$ with a step size $0.1$.
We use a publicly available implementation\footnote{https://github.com/tensorflow/privacy} of DP-SGD. We tried the parameter $noise\_multiplier$ that controls the privacy budget in the range $[0,0.2]$ with a step size $0.05$.
We tried $[0,0.1,0.3,0.5,0.7,1.0]$ as the $\epsilon$ in MemGuard and MemGuard-Random.

\begin{table}
	\centering
	\caption{Results of Model Stacking.}
	\begin{tabular}{|c|c|c|c|} 
	\hline
			 & Location & Texas100 & CH-MNIST  \\ 
			 \hline
	Inference Acc. &   50.0\%      &     50.8\%    &   50.0\%      \\ 
	\hline
	Average Distortion &   1.63     &   1.28     &    0.81     \\
	\hline
	Label Loss &   56.3\%      &    37.9\%    &   18.3\%      \\ 
	\hline
	\end{tabular}
	\label{accuracy_of_target_model}
	\end{table}

\begin{figure*}[!t]
	 \centering
	 \subfloat[Location]{\includegraphics[width=0.330\textwidth]{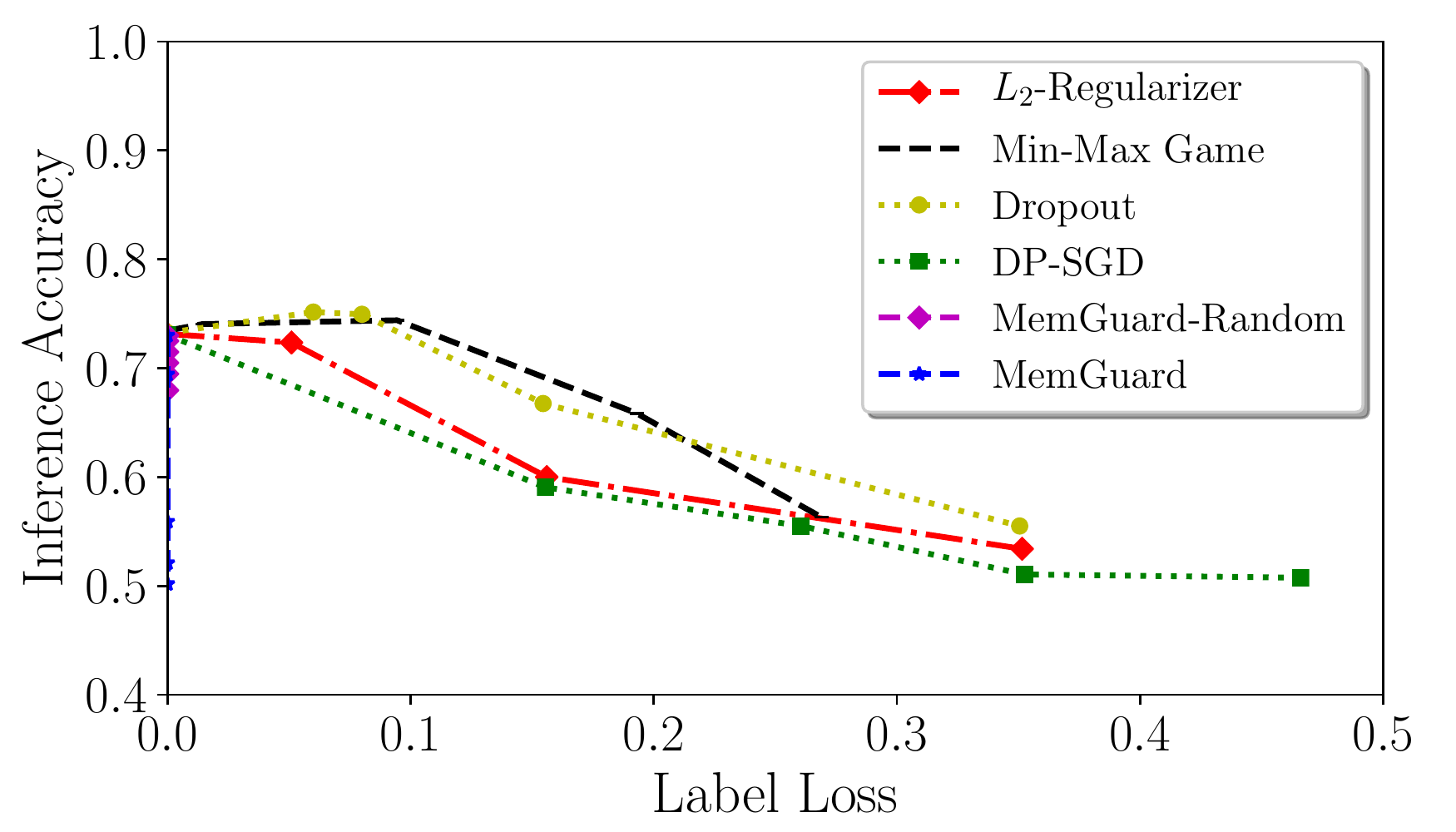}}
	 \subfloat[Texas100]{\includegraphics[width=0.330\textwidth]{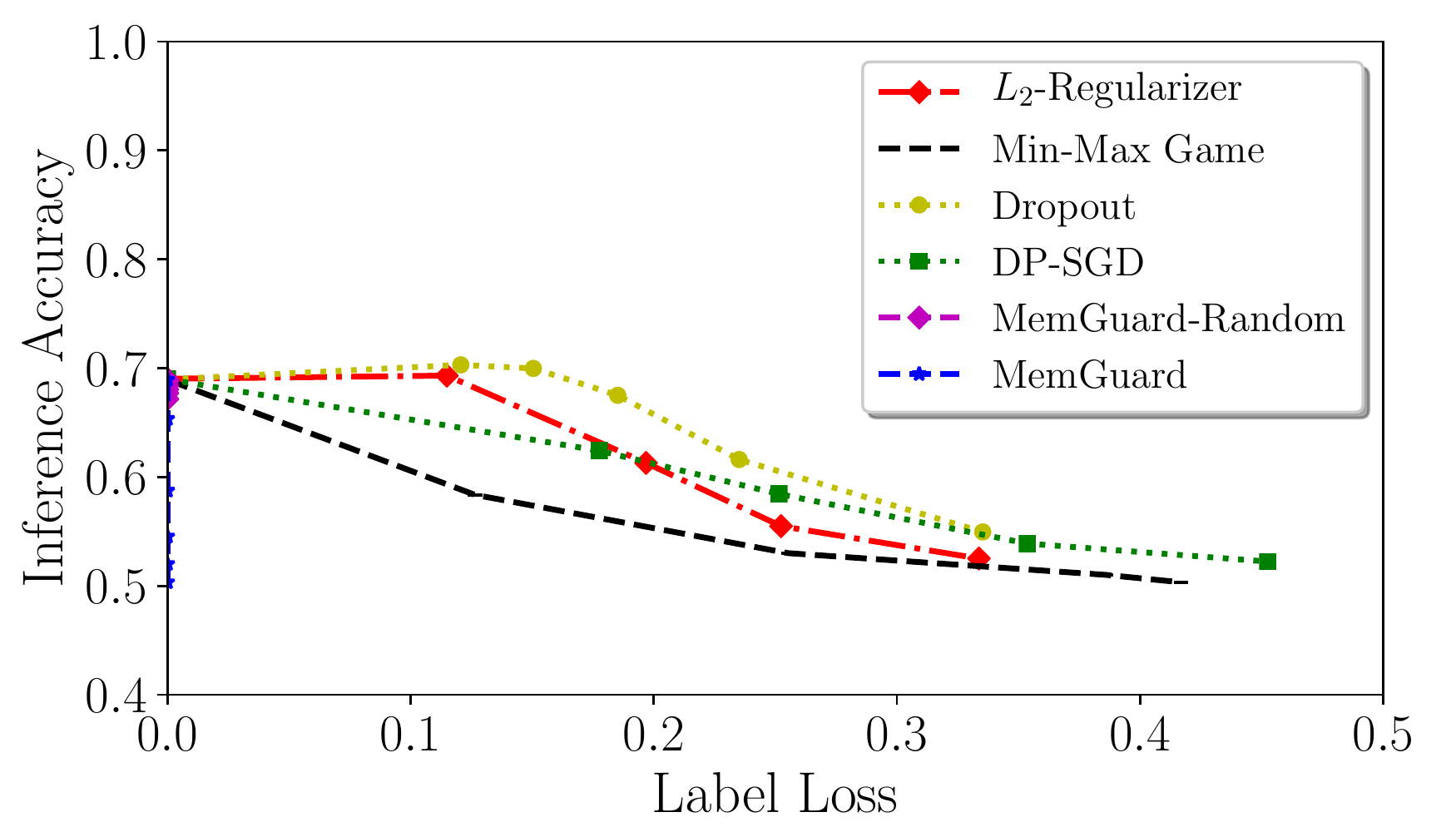}}
	 \subfloat[CH-MNIST]{\includegraphics[width=0.330\textwidth]{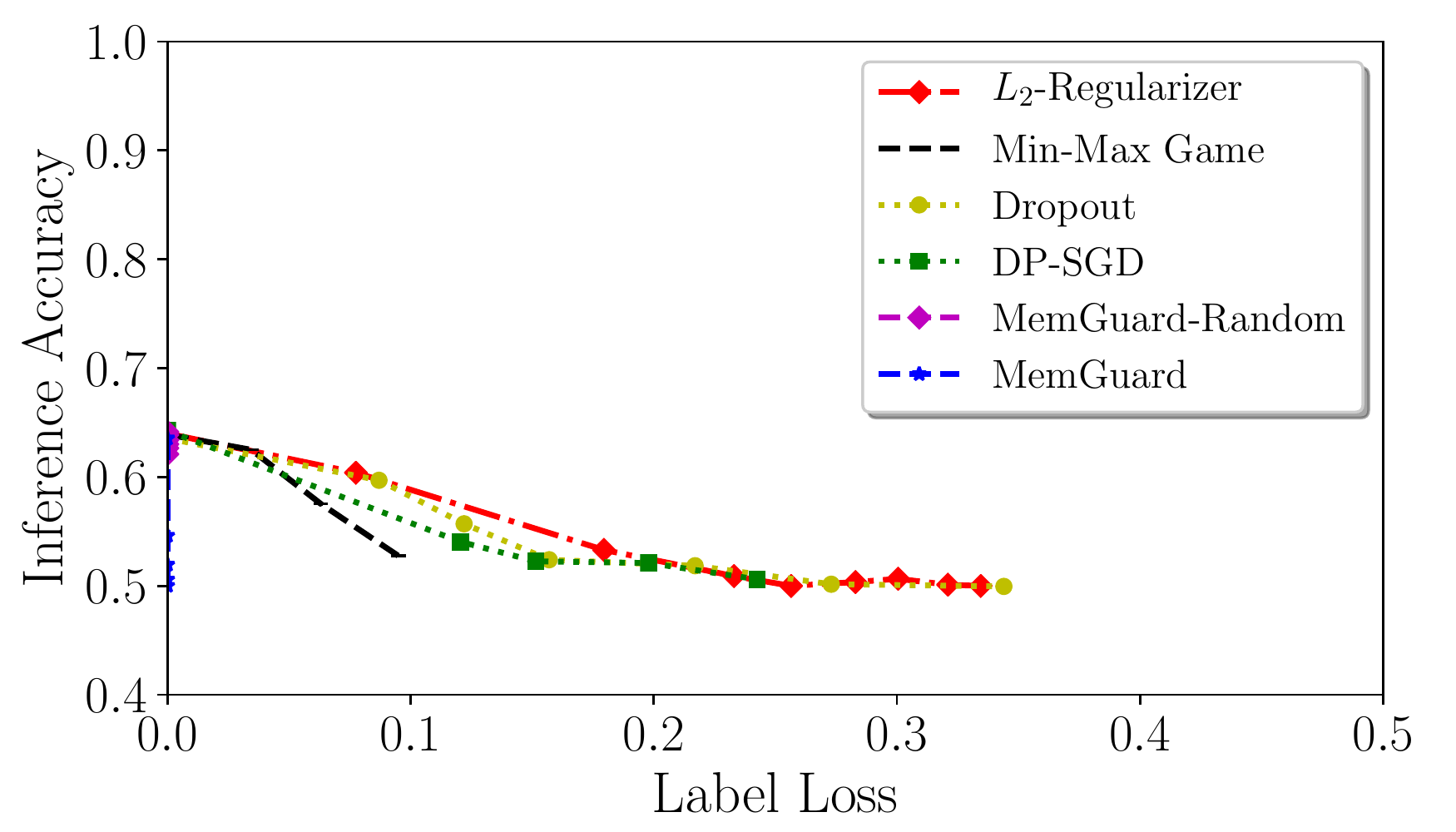}}
	 \caption{Inference accuracy vs. label loss  of the compared defenses. Both MemGuard-Random and MemGuard achieve 0 label loss, while the other defenses incur large label losses in order to substantially reduce the attacker's inference accuracy.}
	 \label{labellossinfer}
 \vspace{-1mm}
\end{figure*}

Our results show that MemGuard achieves the best privacy-utility tradeoff. In particular, given the same average confidence score distortion, MemGuard achieves the smallest inference accuracy. According to the authors of Model Stacking, it does not have a hyperparameter to easily control the privacy-utility tradeoff. Therefore, we just obtain one pair of (inference accuracy, average confidence score distortion) and~\autoref{accuracy_of_target_model} shows the results. Model Stacking reduces the inference accuracy to be close to 0.5, but the utility loss is intolerable.

Similarly, we can obtain a set of pairs (inference accuracy, label loss) for the compared defenses and~\autoref{labellossinfer} shows inference accuracy vs. label loss on the three datasets. Label loss is the fraction of data samples in the evaluation dataset whose predicted labels are changed by a defense.  MemGuard-Random and MemGuard achieve 0 label loss. However, other defenses incur large label losses in order to substantially reduce the attacker's inference accuracy.

%% file: discussion.tex
\section{Discussion and Limitations}
\label{discussion}
On one hand, machine learning can be used by attackers to perform automated inference attacks. On the other hand, machine learning has various vulnerabilities, e.g., \emph{adversarial examples}~\cite{CW17,PMJFCS16,PMGJCS17,SZSBEGF13,PMSW18,PMG16,KGB16,GSS15}. Therefore, attackers who rely on machine learning also share its vulnerabilities and we can exploit such vulnerabilities to defend against them. For instance, we can leverage adversarial examples to mislead attackers who use machine learning classifiers to perform automated inference attacks~\cite{JG19}. One key challenge in this research direction is how to extend existing adversarial example methods to address the unique challenges of privacy protection. For instance, how to achieve formal utility-loss guarantees. 

In this work, we focus on membership inference attacks under the black-box setting, in which an attacker uses a binary classifier to predict a data sample to be a member or non-member of a target classifier's training dataset. In particular, the attacker's classifier takes a data sample's confidence score vector predicted by the target classifier as an input and predicts member or non-member. Our defense adds carefully crafted noise to a confidence score vector to turn it into an adversarial example, such that the attacker's classifier is likely to predict member or non-member incorrectly. To address the challenges of achieving formal utility-loss guarantees, e.g., 0 label loss and bounded confidence score distortion, we design new methods to find adversarial examples. 

Other than membership inference attacks, many other attacks rely on machine learning classifiers, e.g., \emph{attribute inference attacks}~\cite{CAK12,GL162,JWZG17}, \emph{website fingerprinting attacks}~\cite{CZJJ12,JAADG14,WCNJG14,PNZE11,HWF09}, \emph{side-channel attacks}~\cite{ZJRR12}, \emph{location attacks}~\cite{BHPZ17,OTP17,PTC18,ZHRLPB18}, and \emph{author identification attacks}~\cite{NPGBSSS12,CYDHRGN18}. For instance, online social network users are vulnerable to {attribute inference attacks}, in which an attacker leverages a machine learning classifier to infer users' private attributes (e.g., gender, political view, and sexual orientation) using their public data (e.g., page likes) on social networks. The Facebook data privacy scandal in 2018\footnote{\url{https://bit.ly/2IDchsx}} is a notable example of attribute inference attack. In particular, Cambridge Analytica leveraged a machine learning classifier to automatically infer a large amount of Facebook users' various private attributes using their public page likes. Jia and Gong proposed AttriGuard~\cite{JG18}, which leverages adversarial examples to defend against attribute inference attacks. In particular, AttriGuard extends an existing adversarial example method to incorporate the unique challenges of privacy protection. The key difference between MemGuard and AttriGuard is that finding adversarial examples for confidence score vectors is subject to unique constraints, e.g., an adversarial confidence score vector should still be a probability distribution and the predicted label should not change.  Such unique constraints require substantially different methods to find adversarial confidence score vectors. 
Other studies have leveraged adversarial examples to defend against traffic analysis~\cite{ZHRZ19} and author identification~\cite{QMR19,MMJ18}. However, these studies did not consider formal utility-loss guarantees.

We believe it is valuable future work to extend MemGuard to defend against other machine learning based inference attacks such as website fingerprinting attacks, side-channel attacks, and membership inference attacks in the white-box setting. Again, a key challenge is how to achieve formal utility-loss guarantees with respect to certain reasonable utility-loss metrics.

Our MemGuard has a parameter $\epsilon$, which controls a tradeoff between membership privacy and  confidence score vector distortion. The setting of $\epsilon$ may be dataset-dependent. One way to set $\epsilon$ is to leverage an inference accuracy vs. $\epsilon$ curve as shown in Figure~\ref{fix_k_query_infer_acc}. Specifically, given a dataset, we draw the inference accuracy vs. $\epsilon$ curves for various attack classifiers. Suppose we desire the inference accuracy to be less than a threshold. Then, we select the smallest $\epsilon$ such that the inference accuracies of all the evaluated attack classifiers are no larger than the threshold. 

%% file: conclusion.tex
\section{Conclusion and Future Work}
In this work, we propose MemGuard to defend against black-box membership inference attacks. MemGuard is the first defense that has formal utility-loss guarantees on the confidence score vectors predicted by the target classifier. MemGuard works in two phases. In Phase I, MemGuard leverages a new algorithm to find a carefully crafted noise vector to turn a confidence score vector into an adversarial example. The new algorithm considers the unique utility-loss constraints on the noise vector. In Phase II, MemGuard adds the noise vector to the confidence score vector with a certain probability, for which we derive an analytical solution. Our empirical evaluation results show that MemGuard can effectively defend against black-box membership inference attacks and outperforms existing defenses. 
An interesting future work is to extend MemGuard to defend against other types of machine learning based inference attacks such as white-box membership inference attacks, website fingerprinting attacks, and side-channel attacks. 


%% file: appendix.tex
\appendix

\section{Synthesizing Non-members}
\label{synthesizenonmembers}

\begin{figure}[!h]
	 \centering
	 \includegraphics[width=0.33\textwidth]{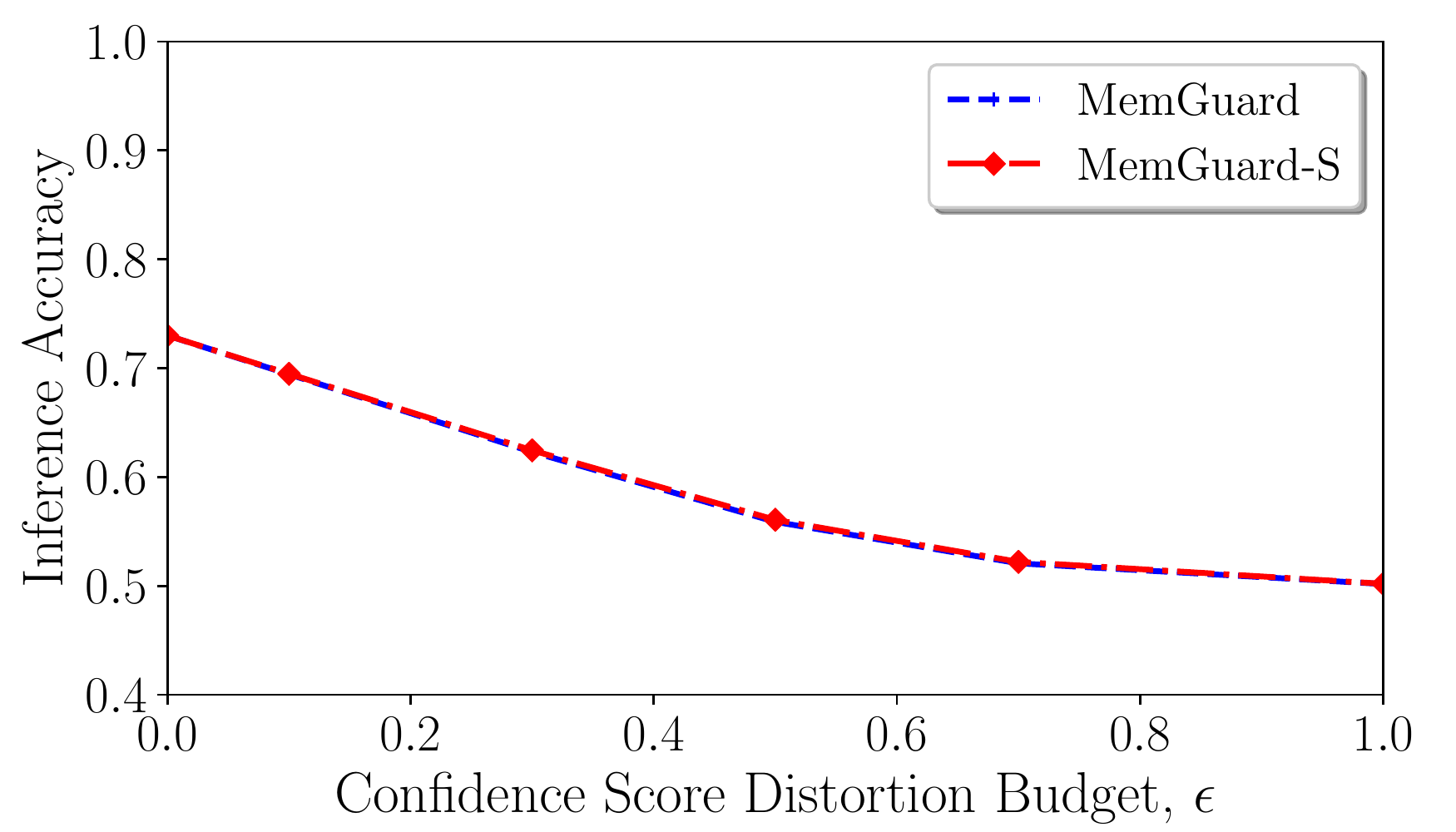}\\
	 \caption{Inference accuracy of the NN attack as the confidence score distortion budget increases on the Location dataset when synthesizing non-members for training the defense classifier (MemGuard-S).}
	 \label{fig:synthesize-app}
\end{figure}

When training the defense classifier, we can use $D_1$ as members and synthesize non-members based on $D_1$. For instance, for each data sample in $D_1$ and each of its feature, we keep the feature value with a probability 0.9 and randomly sample a value from the corresponding data domain for the feature with a probability 0.1, which synthesizes a non-member data sample. Then, we train the defense classifier using $D_1$ as members and the synthesized data samples as non-members. 
\autoref{fig:synthesize-app} 
shows the comparison results on the Location dataset (binary features), where MemGuard-S is the scenario where we synthesize the non-members for training the defense classifier. We observe that MemGuard and MemGuard-S achieve similar performance. Our results show that MemGuard does not necessarily need to split the training dataset in order to train the defense classifier.